\documentclass[%
aip,amsmath,amssymb,reprint
]{revtex4-2}

\usepackage{graphicx}% Include figure files
\graphicspath{{./fig/}}
\usepackage{dcolumn}% Align table columns on decimal point
\usepackage{bm}% bold math
\usepackage[utf8]{inputenc}
\usepackage{mathptmx}
\usepackage{etoolbox}
\usepackage[abs]{overpic}
\usepackage{newtxtext}
\usepackage{physics}
\usepackage{amsmath,mathtools}
\usepackage[colorlinks=true, linkcolor=blue, citecolor=blue, urlcolor=blue]{hyperref}

\makeatletter
\def\@email#1#2{%
 \endgroup
 \patchcmd{\titleblock@produce}
  {\frontmatter@RRAPformat}
  {\frontmatter@RRAPformat{\produce@RRAP{*#1\href{mailto:#2}{#2}}}\frontmatter@RRAPformat}
  {}{}
}%
\makeatother

\begin{document}

%\preprint{APS/123-QED}

\title{Scaling laws for velocity profile of granular flow in rotating drums}

\author{Hiroki Oba}
 \email{hoba.ou.backup@gmail.com}
\affiliation{%
Graduate School of Engineering Science, Osaka University, 1-3 Machikaneyama, Toyonaka, Osaka, 560-8531, Japan }%

\author{Michio Otsuki}%
 \email{otsuki@riko.shimane-u.ac.jp}
\affiliation{%
Institute of Science and Engineering, Shimane University, 1060 Nishikawatsu-cho, Matsue, Shimane, 690--8504, Japan }%

\date{\today}% It is always \today, today,

\begin{abstract}
  We theoretically and numerically investigate the steady flow of two-dimensional granular materials in a rotating drum using the discrete element method and a continuum model with the $\mu(I)$-rheology.
The velocity fields obtained from both methods are in quantitative agreement.
The granular flow exhibits two distinct regions: a surface flow layer and a static flow regime corresponding to rigid rotation near the drum bottom.
The thickness of the surface flow layer increases with the drum diameter and shows a weak dependence on the angular velocity of the drum.
Using dimensional analysis of the continuum equations, we analytically identify nondimensional parameters for the velocity profile and the surface flow layer thickness, which lead to scaling laws characterising the flow in rotating drums with low Froude number and large system size.
The validity of the scaling laws is confirmed by numerical simulations.
\end{abstract}

\maketitle

%\tableofcontents

\section{Introduction}
\label{sec:Introduction}

\begin{figure}
\begin{center}
    \includegraphics[width=1.0\linewidth]{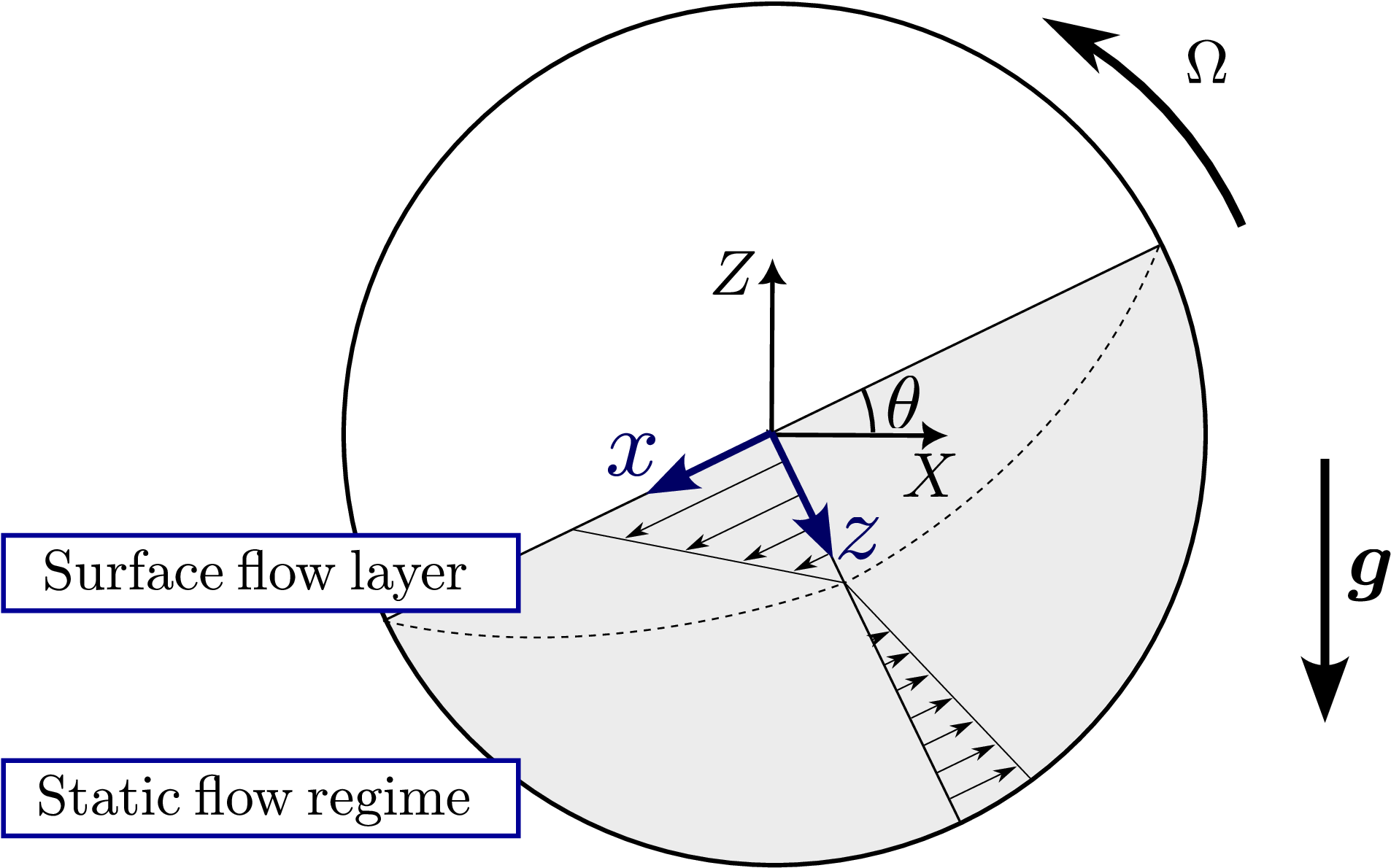}
\end{center}
    \caption{
    Schematic of steady granular flow in a rotating drum.
$X$ and $Z$ denote the horizontal and vertical coordinates, respectively, with the origin located at the drum center.
The $x$-axis is taken parallel to the free surface, and the $z$-axis is normal to it.
    }
    \label{fig:model}
\end{figure}

Granular flows in rotating drums play a central role in a wide range of industrial processes, such as mixing, size reduction, granulation, drying, and chemical reactions.
The flow regime inside the drum is primarily characterized by the Froude number, $Fr=\Omega^{2}D/2g$, where $\Omega$ is the angular velocity, $D$ is the drum diameter, and $g$ is the gravitational acceleration. \cite{Mellmann2001}
For the {\it rolling} state with $10^{-4}<Fr<10^{-2}$, the system exhibits a steady flow \cite{Mellmann2001}, which can be divided into two distinct regions.
Near the free surface, a thin ``surface flow layer'' is formed where particles continuously cascade down the slope.
Beneath this layer, particles undergo a solid-body rotation with the drum wall, often referred to as the ``static flow regime'' (see Fig. \ref{fig:model}).

Numerous experimental studies have examined how the thickness of the surface flow layer, $h$, in the {\it rolling} state depends on the drum diameter $D$ and angular velocity $\Omega$ in half-filled drums \cite{Parker1997,Khakhar1997,Jain2004,Felix2007,Orpe2007,Chen2024}.
For example, Parker et al. \cite{Parker1997} studied axially long drums with glass beads and reported that $h$ scales linearly with $D$ and is nearly independent of $\Omega$.
In contrast, F\'{e}lix et al. \cite{Felix2007} investigated thin drums and found that $h$ measured near the walls exhibits a power-law dependence on $\Omega$ when $D \gg d$.
Pignatel et al. \cite{Pignatel2012} further suggested that $h/d$ for glass and steel spheres in thin drums sandwiched between acrylic walls follows $h/d \propto \tilde{Q}^{0.44}$, where $\tilde{Q} = \tfrac{1}{2} Fr^{1/2} (D/d)^{3/2}$ is the dimensionless flow rate.
Taken together, these studies indicate that the scaling of $h$ depends on the system geometry, sidewall effects, and material properties, and varies significantly across different Froude and inertial number regimes.
However, a consistent theoretical explanation for these dependencies has not yet been established.

Granular flows in rotating drums have also been studied extensively using discrete element method (DEM) simulations, in which the equations of motion for individual particles are solved numerically. \cite{Yamane1997,Dury1998,Renouf2005,Yang2003,Yang2008} 
These simulations have successfully reproduced the velocity profiles observed in experiments. 
However, direct simulations of systems at industrially relevant scales remain impractical, since the computational cost of DEM grows rapidly with the particle number when $D \gg d$. 
Moreover, while DEM provides detailed particle-level information, it does not by itself yield a theoretical description of the dependence of the surface flow layer thickness $h$ on $D$ and $\Omega$.

Another approach is to use continuum equations to investigate granular flows in rotating drums, providing a macroscopic description of the flow behavior.
Among various constitutive relations proposed in this framework, the so-called $\mu(I)$-rheology has become widely used due to its simplicity and ability to capture dense granular flows. \cite{DaCruiz2005,Jop2006,Peyneau2008} 
%This rheology effectively treats granular media as a non-Newtonian fluid whose viscosity depends on the local shear rate and pressure, and extensions incorporating non-local effects have also been developed. \cite{Kamrin2012,Heyman2017,Kim2020} 
In the $\mu(I)$-rheology, granular media is treated as a non-Newtonian fluid whose viscosity depends on the local shear rate and pressure, and extensions incorporating non-local effects have also been developed. \cite{Kamrin2012,Heyman2017,Kim2020} 
Continuum models based on such constitutive relations have successfully reproduced velocity profiles in rotating drums across several setups. \cite{G.D.R.MiDi2004,Zheng2015,Zheng2019,Liu2018,Arseni2020,Santos2013} 
Nevertheless, systematic studies on how these models capture the dependence of the surface flow layer thickness $h$ on $D$ and $\Omega$ are still lacking. 
In addition, most prior work has relied primarily on computational fluid dynamics (CFD) simulations of the continuum model to compute the velocity field numerically, rather than on analytical understanding.

For conventional fluids, theoretical understanding has been achieved by analyzing the continuum equations, i.e., the Navier-Stokes equations. \cite{Batchelor2000}
If continuum models for granular materials can successfully reproduce experimental and DEM velocity profiles, one may expect that similar theoretical analyses would clarify the dependence of the velocity field on system parameters such as $D$ and $\Omega$ in rotating drums.
However, to the best of our knowledge, such a theoretical explanation has not yet been developed, despite the numerical evidence of the validity of continuum descriptions for granular flows.

This study aims to provide a theoretical understanding of steady granular flows in rotating drums.
As a first step, we focus on two-dimensional drums half-filled with granular materials in the limit $D/d \gg 1$.
This setting enables an analytically tractable formulation of the continuum model and remains directly relevant to experiments in sufficiently long drums, where the flow can be approximated as two-dimensional.
We first demonstrate, by comparing with discrete element method (DEM) simulations, that the continuum model with the $\mu(I)$-rheology quantitatively reproduces the velocity field.
Building on this validation, we then derive scaling laws for the velocity profile and the surface flow layer thickness $h$ from the continuum model and verify them against DEM results.

This paper is organized as follows.
Section \ref{sec:NumericalMethod} describes the setup of our models.
Section \ref{sec:Results} presents the velocity fields obtained from DEM and continuum simulations.
Scaling laws derived from the continuum model are given in Sec. \ref{sec:ScalingLaws}.
Sec. \ref{sec:DiscussionAndConclusion} summarizes the main findings and discusses possible directions for future work.
Additional details, including the determination of rheological parameters and analytical derivations, are provided in the Appendices.

%----------------------------------------------%
% Numerical Method                             %
%----------------------------------------------%
\section{Setup}
\label{sec:NumericalMethod}

In this study, we consider a two-dimensional granular flow in a rotating drum, as illustrated in Fig.~\ref{fig:model}.  
This choice captures the essential features of the flow relevant to testing the theoretical predictions, while keeping the computational cost manageable.
The origin of the Cartesian coordinate system $(X, Z)$ is placed at the center of the drum with diameter $D$.
Polydisperse particles with minimum diameter $d$ are half-filled in the drum.  
The drum rotates counterclockwise with angular velocity $\Omega$.
We consider three drum diameters: $D = 150d$, $300d$, and $450d$.
The angular velocity is varied within $1.26 \times 10^{-3} \le \Omega/\sqrt{g/d} \le 1.26 \times 10^{-2}$ to satisfy $10^{-4} < Fr < 10^{-2}$.  
To provide reference data for the continuum analysis, we first perform discrete element method (DEM) simulations in this 2D system.  
The particle-scale information obtained from DEM serves as a benchmark for validating the continuum model, which will be introduced in Sec. \ref{subsec:CFD} and serves as the primary tool for deriving theoretical insights into the velocity field.

\subsection{DEM simulation}
\label{subsec:DEM}

In the DEM simulation, \cite{Cundall1979,Shafer1996,Campbell2002} we consider $N$ particles of identical density $\rho_s$ in a drum.  
Particle interactions are modeled as linear springs, characterized by the normal elastic constant $k^{(\mathrm{n})}$, tangential elastic constant $k^{(\mathrm{t})}$, restitution coefficient $e$, and particle friction coefficient $\mu_{\mathrm{p}}$.  
The rough wall surface of the drum is modeled by placing particles of diameter $d$.

We use polydisperse particles, \cite{Renouf2005} with diameters uniformly distributed from $d$ to $1.2d$.  
The parameters are set as follows: $k^{(\mathrm{n})}=2.0 \times 10^5 mg/d$, $k^{(\mathrm{t})}/k^{(\mathrm{n})}=2/7$, $e=0.92$, and $\mu_p=0.50$, \cite{Silbert2001} where $m$ is the mass of a particle of diameter $d$.  
The time evolution equations are solved numerically using the leapfrog algorithm with time step $\Delta t = 1.0 \times 10^{-4} \sqrt{d/g}$.
The drum is half-filled with $N$ particles, set to $7700$, $30600$, and $69200$ for $D=150d$, $300d$, and $450d$, respectively.  
Details of the DEM simulation are given in Appendix \ref{App:DetailOfDEM}.  
From the DEM simulation, we calculate the velocity field in the steady state using the procedure described in Appendix \ref{appA}.

\subsection{Continuum model}
\label{subsec:CFD}

To gain a theoretical understanding of the flow behavior observed in the DEM simulations, 
we adopt a continuum model that treats the granular bed in the rotating drum as an incompressible, non-Newtonian fluid. 
The governing equations consist of the mass and momentum conservation equations.
The mass conservation equation is given by
\begin{equation}
    \nabla \cdot \vb*{v} = 0,
    \label{eq:Mass}
\end{equation}
and the momentum conservation equation reads
\begin{equation}
    \frac{\partial \vb*{v}}{\partial t} = -  \left( \vb*{v} \cdot \nabla \right)  \vb*{v} - \frac{1}{\rho_{\rm g}} \nabla p + \frac{1}{\rho_{\rm g}} \nabla \cdot \vb*{\tau} + \vb*{g},
    \label{eq:Momentum}
\end{equation}
with $\vb*{g} = (0,-g)$,
where $\rho_{\rm g}$, $\vb*{v}$, and $p$ denote the granular density, velocity, and pressure at position $\vb*{r} = (X,Z)$ and time $t$, respectively.
%To close the system, we specify the constitutive relation for the stress tensor $\vb*{\tau}$ in the following.

In this study, we adopt the $\mu(I)$ rheology \cite{DaCruiz2005,Jop2006} as the constitutive relation for the stress tensor $\vb*{\tau}$, which is expressed as
\begin{equation}
    \vb*{\tau} = \mu(I)\, p\, \frac{ \vb*{\dot{\gamma}} }{ \dot{\gamma} },
    \label{eq:Constitutive}
\end{equation}
where the shear-rate tensor is defined by $\vb*{\dot{\gamma}} = \nabla \vb*{v} + (\nabla \vb*{v})^T$.
The effective friction coefficient $\mu(I)$ depends on the inertial number $I = \dot{\gamma} / \sqrt{p / \rho_s d^2}$, 
where $\dot{\gamma}$ denotes the second invariant of $\vb*{\dot{\gamma}}$ and $\rho_s$ is the material density of the particles.
Following Ref. \onlinecite{Jop2006}, $\mu(I)$ is given by
\begin{equation}
    \mu(I) = \mu_{s} + \frac{\mu_{2} - \mu_{s}}{I_0/I + 1},
    \label{eq:mu_I_Rheology_Function}
\end{equation}
where $\mu_s$, $\mu_2$, and $I_0$ are material parameters reflecting the elasticity, dissipation, and friction of the particles. \cite{DaCruiz2005}
At constant pressure $p$, this relation reduces to the Bingham-type model, 
which exhibits a finite yield stress below which no flow occurs.
In this framework, granular materials can be regarded as non-Newtonian fluids 
whose viscosity depends on the local shear rate and pressure as
\begin{equation}
    \eta_{\mathrm{g}}(\dot{\gamma}, p) = \frac{ \mu(I)\, p }{ \dot{\gamma} }.
    \label{eq:Viscosity}
\end{equation}

At the wall of the drum, the velocity satisfies
\begin{equation}
    \vb*{v} = \vb*{v}_{\mathrm{wall}},
    \label{eq:BCWall}
\end{equation}
where
\begin{equation}
    \vb*{v}_{\mathrm{wall}} = -\Omega Z \vb*{e}_X + \Omega X \vb*{e}_Z,
    \label{eq:WallVelocity}
\end{equation}
and $\vb*{e}_X$ and $\vb*{e}_Z$ are the unit vectors along the $X$ and $Z$ axes, respectively. 
At the free surface of the granular bed, we impose the stress-free boundary conditions:
\begin{equation}
    \vb*{\tau} \cdot \vb*{n} = \vb*{0},
    \label{eq:BCFree2}
\end{equation}
where $\vb*{n}$ denotes the unit normal vector to the free surface.

We employ the continuum model to theoretically derive the scaling law for the velocity profile, after confirming numerically that the model quantitatively reproduces the results of the DEM simulation.
To compute the velocity field based on the continuum equations, we employ CFD with the finite-difference method, using a time step $\Delta t$ and grid width $\Delta x$, following Ref. \onlinecite{Watanabe2022}, which studied an incompressible Newtonian fluid in a rotating container.
Details of the CFD simulations are provided in Appendix \ref{App:CFD}.
The parameters $\mu_{s}$, $\mu_{2}$, and $I_{0}$ in Eq. \eqref{eq:mu_I_Rheology_Function} are estimated as $\mu_{s} = 0.246$, $\mu_{2} = 0.401$, and $I_{0} = 0.133$ from the DEM simulation under uniform shear with the parameters given in Sec. \ref{subsec:DEM}. 
The details of this estimation procedure are shown in Appendix \ref{appB}.
Note that Eq. \eqref{eq:mu_I_Rheology_Function} is also used as the constitutive relation for grains interacting via the Hertzian contact model, with appropriate choices of these parameters. \cite{DaCruiz2005}
The granular density is defined as $\rho_{\mathrm{g}} = \rho_s \phi$, where $\rho_s = m / (\pi d^2 / 4)$ is the material density.
The packing fraction $\phi$ is set to $0.825$, corresponding to its value in the static regime of the rotating drum obtained from the DEM simulation (Appendix \ref{appD}).
The spatial and temporal resolutions are set to $\Delta x = D/75$ and $\Delta t = 1.0 \times 10^{-4} \sqrt{d/g}$, respectively.

%----------------------------------------------%
% Results                                      %
%----------------------------------------------%
\section{Validation of the Continuum Model}
\label{sec:Results}

\begin{figure}
    \centering
    \includegraphics[width=0.9\linewidth]{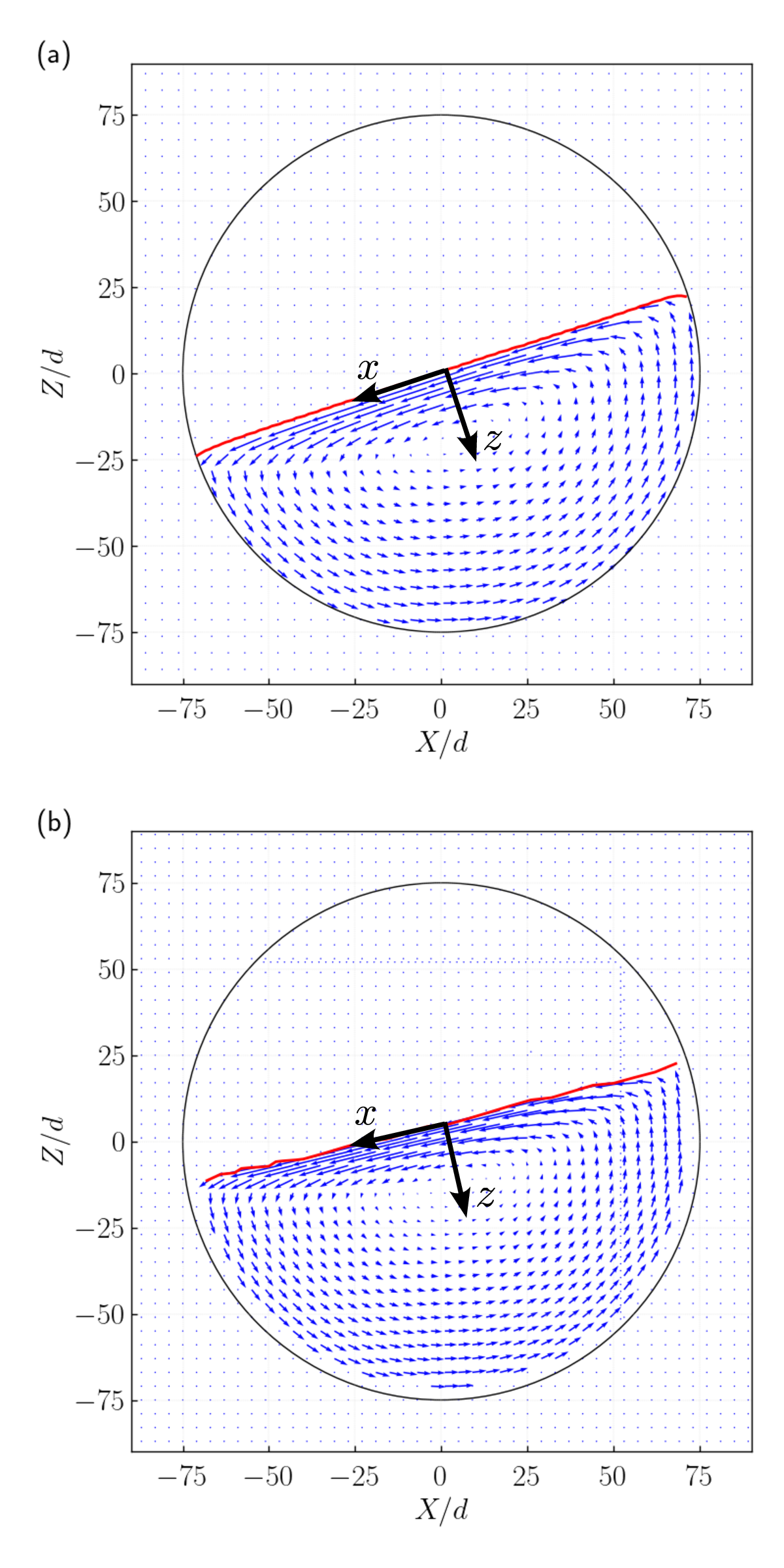}
    \caption{
      Velocity fields obtained from the DEM (a) and CFD (b) simulations for $D=150d$ and $\Omega = 1.26 \times 10^{-3} \sqrt{g/d}$.
The red solid line represents the free surface.
The $x$-axis is taken parallel to the free surface, and the $z$-axis is normal to it.
    }
    \label{fig:velocity_vmap}
\end{figure}

To assess the validity of the continuum model introduced in Sec. \ref{subsec:CFD}, we compare the velocity field in the steady state obtained from DEM simulations and from CFD calculations that numerically solve the continuum equations.
Figures \ref{fig:velocity_vmap} (a) and (b) show the velocity field of granular material for $D=150d$ and $\Omega = 6.28 \times 10^{-3} \sqrt{g/d}$ obtained from the DEM and CFD simulations, respectively.
The red solid lines in these figures represent the free surface.
In both cases, the granular material convects in a counterclockwise direction.
Near the wall, the flow exhibits the rigid rotation following the rotating drum.
Near the free surface, there is a region where the velocity is almost parallel to the free surface.

To facilitate a quantitative comparison of the velocity field between DEM and CFD, we introduce a new Cartesian coordinate system $(x,z)$, where the $x$ axis is parallel to the free surface at $X=0$, as shown in Fig. \ref{fig:model}.
In this coordinate system, the depth from the free surface is represented by $z$.
The coordinate transformation from $(X,Z)$ to $(x,z)$ is given by
\begin{equation}
    \left(
    \begin{array}{c}
    x \\
    z
    \end{array}
    \right)
    =
    \left(
    \begin{array}{cc}
    - \cos{\theta} & - \sin{\theta}\\
    \sin{\theta} & - \cos{\theta}
    \end{array}
    \right)
    \left(
    \begin{array}{c}
    X \\
    Z
    \end{array}
    \right),
\end{equation}
where $\theta$ denotes the angle of the free surface at $X=0$.
The position of the free surface in the DEM simulation is determined from the packing fraction, as detailed in Appendix \ref{appD}, and the free surface at $x=0$ is treated as $z = 0$, which is numerically evaluated in Appendix \ref{appE}.

\begin{figure}
    \centering
    \includegraphics[width=1.0\linewidth]{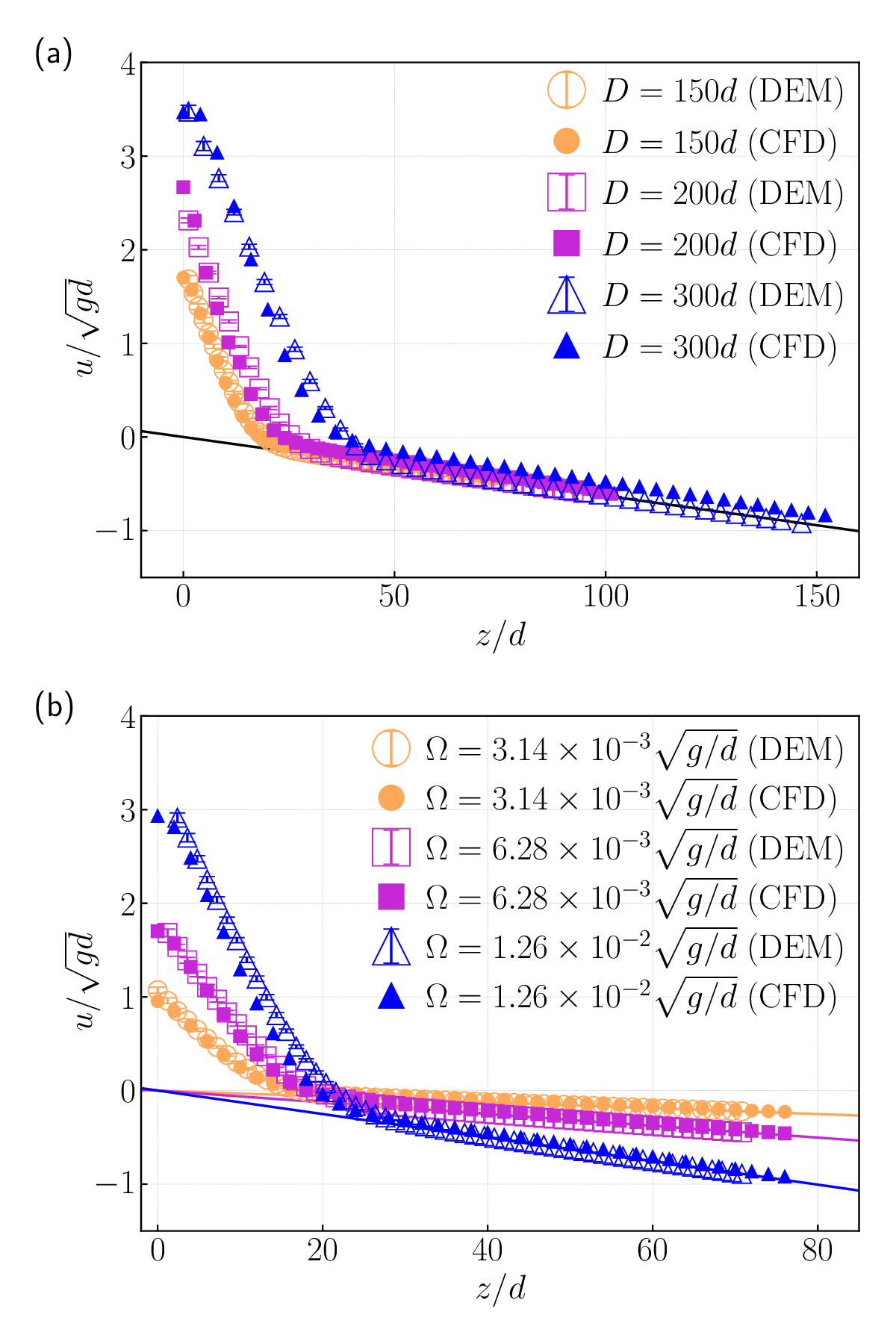}
    \caption{(a) Velocity profile $u(z)$ at $x=0$ for $\Omega = 6.28 \times 10^{-3} \sqrt{g/d}$ with $D=150d$, $200d$, and $300d$. (b) Velocity profile $u(z)$ at $x=0$ for $D=150d$ with $\Omega = 1.26 \times 10^{-3} \sqrt{g/d}$ to $1.26 \times 10^{-2} \sqrt{g/d}$.
    The open and closed symbols represent the results of the DEM and CFD simulations, respectively.
    The solid lines represent $- \Omega z$ for the rigid rotation.
    }
    \label{fig:velocity_center}
\end{figure}

Next, we analyze the velocity $u$ in the $x$-direction as a function of $z$ at the center of the granular bed ($x=0$).
Figures \ref{fig:velocity_center} show the velocity profiles for various values of $D$ and $\Omega$, obtained from DEM and CFD simulations.
Open and closed symbols represent DEM and CFD results, respectively, which are quantitatively consistent.
The velocity decreases with increasing $z$, approaching the rigid rotation profile $- \Omega z$ at larger $z$, corresponding to the static flow regime, while significant deviations near $z=0$ correspond to the surface flow layer.
The magnitude and slope of $u(z)$ in the surface flow layer increase with $D$ and $\Omega$, and the two regions are smoothly connected at the boundary where $u(z)\approx 0$.
These results agree with Ref. \onlinecite{Renouf2005} and demonstrate that the continuum model using the $\mu(I)$-rheology quantitatively reproduces the velocity field over a broad parameter space, beyond the limited conditions studied previously \cite{G.D.R.MiDi2004,Zheng2015,Zheng2019,Arseni2020}.

%Using this $(x,z)$ coordinate system, we now quantitatively compare the velocity profiles obtained from DEM and CFD simulations.
%In Fig. \ref{fig:velocity_center} (b), the velocity profile $u(z)$ obtained from the DEM and CFD simulations for various values of $\Omega$ with $D=150d$ is shown.
%The results of the DEM and CFD simulations are quantitatively consistent in this figure.
%For each $\Omega$, the surface flow layer and the static flow regime are distinguished.
%The magnitude and the slope of $u(z)$ in the surface flow layer increase with $\Omega$.
%Figure \ref{fig:velocity_center} demonstrates that the continuum model using the $\mu(I)$-rheology quantitatively reproduces the velocity field over a broad parameter space, extending beyond the limited conditions examined in previous studies. \cite{G.D.R.MiDi2004,Zheng2015,Zheng2019,Arseni2020}

\begin{figure}
    \centering
    \includegraphics[width=1.0\linewidth]{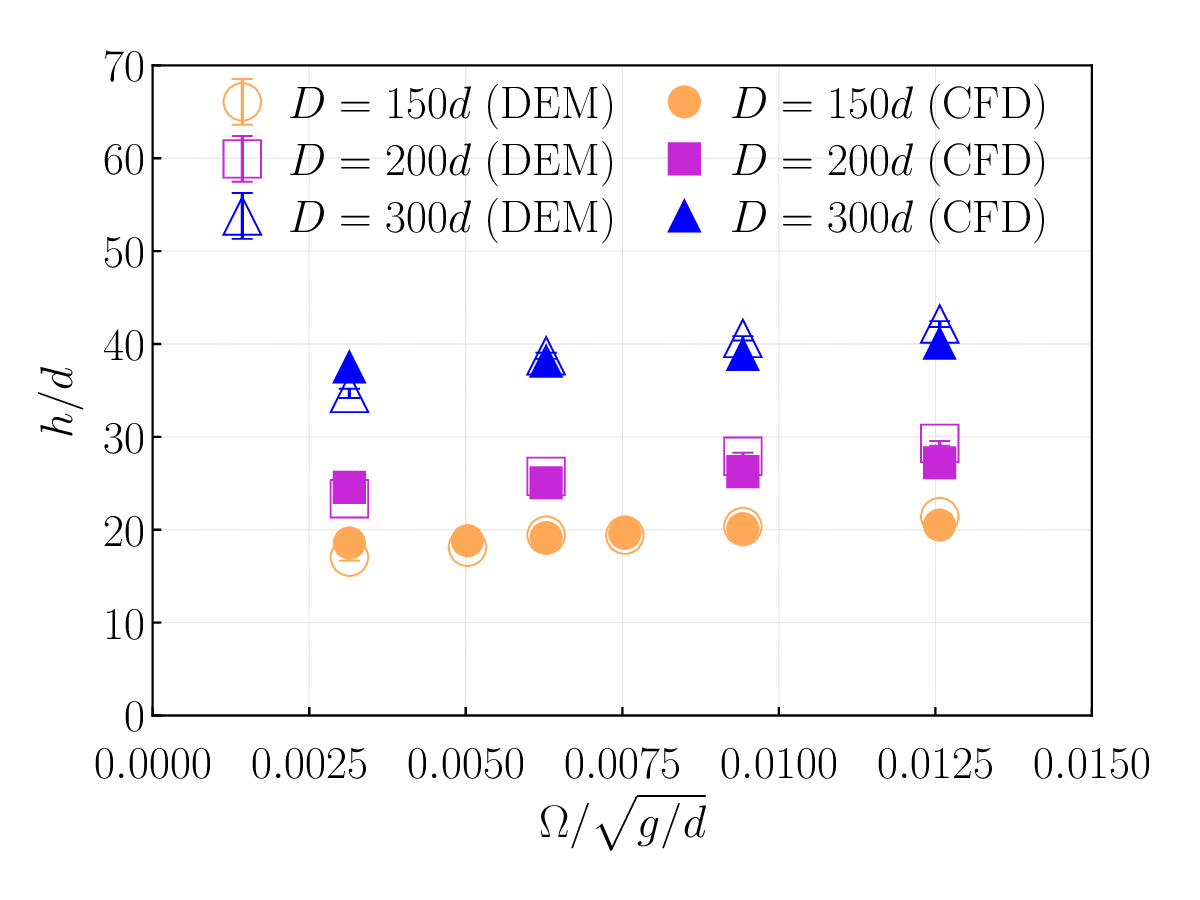}
    \caption{Surface flow layer thickness $h$ against the angular velocity $\Omega$ for various values of $D$.
    The open and closed symbols represent the results of the DEM and CFD simulations, respectively.
    }
    \label{fig:Omega-h}
\end{figure}

Next, to investigate the dependence of the velocity profile on $D$ and $\Omega$, we define the surface flow layer thickness $h$.
Following Refs. \onlinecite{Felix2007,Pignatel2012}, we identify the point where $u(z)=0$ in Fig. \ref{fig:velocity_center} as the boundary between the surface flow layer and the underlying static bed, and estimate the surface flow layer thickness $h$ as the distance from the free surface to this boundary at $x=0$:
\begin{equation}
u(z=h)=0.
\label{eq:u_h_relation}
\end{equation}
Figure \ref{fig:Omega-h} shows the surface flow layer thickness $h$ obtained from the DEM and CFD simulations as a function of $\Omega$ for different $D$.
The numerical results from the DEM and CFD simulations are in good agreement.
The surface flow layer thickness $h$ increases with $D$ and shows a weak dependence on $\Omega$.
Although previous studies have estimated the boundary using different methods, \cite{Ding2002,Bonamy2002,G.D.R.MiDi2004,Renouf2005,Cortet2009} the choice of method does not affect the qualitative behavior of $h$, \cite{Felix2007} which can be explained by the scaling of the velocity profile presented below (Appendix \ref{appF}).

%----------------------------------------------%
% Scaling laws                                 %
%----------------------------------------------%
\section{Scaling laws}
\label{sec:ScalingLaws}

In this section, based on the continuum model introduced in Sec. \ref{subsec:CFD}, we theoretically derive scaling laws for the velocity profile of granular flow in a rotating drum.
We neglect the inertial term $\partial \vb*{v}/\partial t$ in Eq. \eqref{eq:Momentum} by assuming steady flow.
Using the drum diameter $D$, angular velocity $\Omega$, and granular density $\rho_{\mathrm{g}}$ as characteristic scales, we non-dimensionalize the variables as follows:
\begin{align}
  & \tilde{\vb*{r}} = \frac{\vb*{r}}{D}, \;\;
\tilde{\nabla} = D \nabla, \;\;
\tilde{\vb*{v}} = \frac{\vb*{v}}{\Omega D}, \;\; \\
  & \tilde{\dot{\gamma}} = \frac{\dot{\gamma}}{\Omega}, \;\;
\tilde{p} = \frac{p}{\rho_{\mathrm{g}} \Omega^2 D^2}, \;\;
  \tilde{\eta}_{\mathrm{g}} = \frac{\eta_{\mathrm{g}}}{\rho_{\mathrm{g}} \Omega D^2},
\end{align}
where $\tilde{\vb*{r}}, \tilde{\nabla}, \tilde{\vb*{v}}, \tilde{\dot{\gamma}}, \tilde{p}$, and $\tilde{\eta}_{\mathrm{g}}$ are the dimensionless variables.  
We use $D$ as the characteristic length because we focus on the overall velocity profile in systems with $D \gg d$.
Substituting these dimensionless variables into Eqs. \eqref{eq:Mass}--\eqref{eq:Viscosity}, we obtain
\begin{gather}
  \tilde{\nabla} \cdot \tilde{\vb*{v}} = 0 , \label{eq:Mass_normalize}\\
- \tilde{\nabla} \cdot (\tilde{\vb*{v}} \tilde{\vb*{v}}) - \tilde{\nabla} \tilde{p} 
+ \tilde{\nabla} \cdot \Big[ \tilde{\eta}_{\mathrm{g}} \left( \tilde{\nabla} \tilde{\vb*{v}} + (\tilde{\nabla} \tilde{\vb*{v}})^T \right) \Big] 
  - \frac{2}{Fr} \vb*{e}_Z = 0 , \label{eq:Momentum_normalize} \\
\tilde{\eta}_{\mathrm{g}} = \mu_s \frac{\tilde{p}}{\tilde{\dot{\gamma}}} 
  + (\mu_2 - \mu_s) \frac{\tilde{p}}{I_0 \sqrt{\phi \tilde{p}} / (d/D) + \tilde{\dot{\gamma}}}, \label{eq:Viscosity_normalize}
\end{gather}
where the Froude number is defined as $Fr = \Omega^2 D / (2 g)$.
We have summarized the dimensionless variables in Table \ref{tab:nondimensional}.
In the present system, the dimensionless parameters relevant for the velocity profile are $Fr$ and $d/D$.  
Thus, apart from the particle properties $\mu_s$, $\mu_2$, and $I_0$, the dimensionless velocity field $\tilde{\vb*{v}}(\vb*{r})$ is determined solely by these two parameters.  
This scaling is consistent with the experiments reported in Ref. \onlinecite{Orpe2001}.

\begin{table}[t]
\caption{Summary of dimensionless variables used in this study.}
\label{tab:nondimensional}
\begin{ruledtabular}
\begin{tabular}{ll}
Dimensionless variables & Definition / Physical meaning \\
\hline
$\tilde{\mathbf r} = \mathbf r / D$ & Scaled position \\
$\tilde{\mathbf v} = \mathbf v / (\Omega D)$ & Scaled velocity \\
$\tilde{p} = p / (\rho_g \Omega^2 D^2)$ & Scaled pressure \\
$\tilde{\dot{\gamma}} = \dot{\gamma} / \Omega$ & Scaled shear rate \\
$\tilde{\eta}_g = \eta_g / (\rho_g \Omega D^2)$ & Scaled granular viscosity \\
$Fr = \Omega^2 D / (2g)$ & Froude number \\
$d/D$ & Particle-to-drum diameter ratio \\
$\mu_s, \mu_2, I_0$ & Rheological parameters in the $\mu(I)$-rheology
\end{tabular}
\end{ruledtabular}
\end{table}

Since in the previous section we have studied systems with $d/D \le 1/150 \ll 1$, we consider the limit $d/D \to 0$ in Eqs. \eqref{eq:Mass_normalize}--\eqref{eq:Viscosity_normalize}.  
In this limit, the only remaining dimensionless parameter in these equations is the Froude number $Fr$.  
Hence, the solution of the continuum equations can be expressed as
\begin{equation}
    \tilde{\vb*{v}} \left( \tilde{\vb*{r}} \right) = \vb*{V} \left( \tilde{\vb*{r}} ; Fr \right),
    \label{eq:velocity_Dd-Fr}
\end{equation}
where $\vb*{V} \left( \tilde{\vb*{r}} ; Fr \right)$ is a function that depends on $Fr$.  
Taking the limit $d/D \to 0$ in Eq. \eqref{eq:Viscosity_normalize} is equivalent to treating the granular material as a simple plastic body with a yield stress proportional to the pressure, an approximation similar to that introduced in Ref. \onlinecite{Agarwal2021}.  
From Eq. \eqref{eq:velocity_Dd-Fr}, the scaling law for the velocity profile $u(z)$ at $x=0$ is obtained as
\begin{equation}
    \frac{u(z)}{\Omega D} = U \left( \frac{z}{D} ; Fr \right),
    \label{eq:u_Fr}
\end{equation}
where $U \left( z/D ; Fr\right)$ is a dimensionless scaling function.

Substituting Eq. \eqref{eq:u_Fr} into the definition of the surface flow layer thickness, Eq. \eqref{eq:u_h_relation}, we have
\begin{equation}
    U \left( \frac{h}{D} ; Fr \right) = 0.
\end{equation}
This relation indicates that the dimensionless surface flow layer thickness $h/D$ is solely determined by the Froude number $Fr$, leading to the scaling law
\begin{equation}
    \frac{h}{D} = H \left( Fr \right),
    \label{eq:h_Fr}
\end{equation}
where $H(Fr)$ is a dimensionless scaling function.  
Thus, once the function $H(Fr)$ is known, the dependence of the surface flow layer thickness on $D$ and $\Omega$ is fully determined by the continuum model.

\begin{figure}
    \centering
    \includegraphics[width=1.0\linewidth]{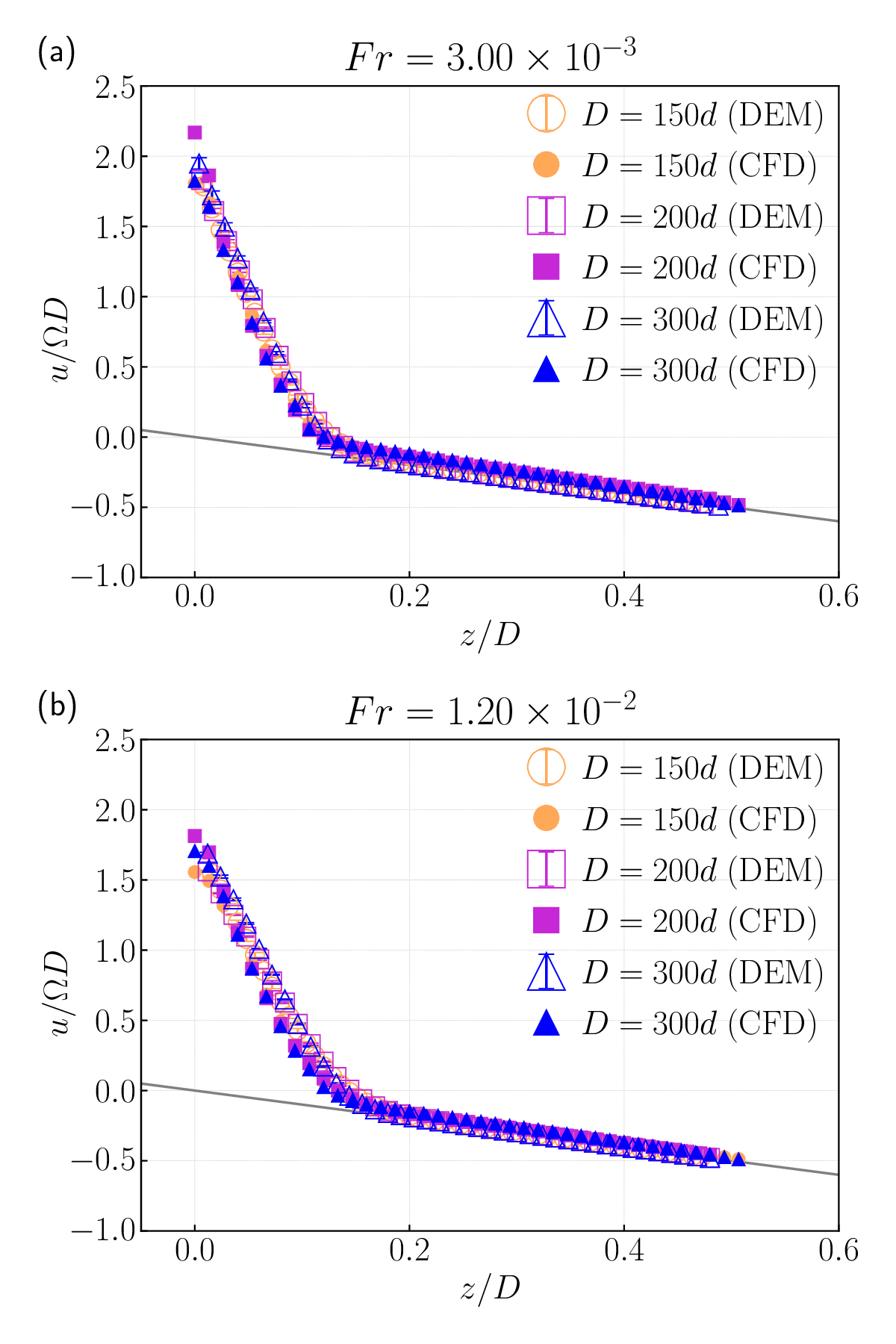}
    \caption{Normalized velocity $u / \Omega D$ against $z/D$ at position $x=0$ for (a) $Fr = 3.00 \times 10^{-3}$, and (b) $Fr = 1.20 \times 10^{-2}$ with various values of $D$.
    The open and closed symbols represent the results of the DEM  and CFD simulations, respectively.
    The solid line corresponds to the rigid rotation.}
    \label{fig:set-Fr_velocity_center}
\end{figure}

We verify the scaling laws using the numerical simulations.
Figure \ref{fig:set-Fr_velocity_center} shows the normalized velocity profile $u / \Omega D$ as a function of $z/D$ for different $D/d$ values, with (a) $Fr = 3.00 \times 10^{-3}$ and (b) $Fr = 1.20 \times 10^{-2}$.
Closed and open symbols represent the results of the CFD and DEM simulations, respectively, while the solid line corresponds to rigid rotation.
Excellent collapse of the data is observed in Fig. \ref{fig:set-Fr_velocity_center}, confirming the validity of the scaling law given in Eq. \eqref{eq:u_Fr}.
Comparing the results for different $Fr$, we also find that the scaling function $U \left( z/D ; Fr\right)$ exhibits a weak dependence on $Fr$.

\begin{figure}
    \centering
    \includegraphics[width=1.0\linewidth]{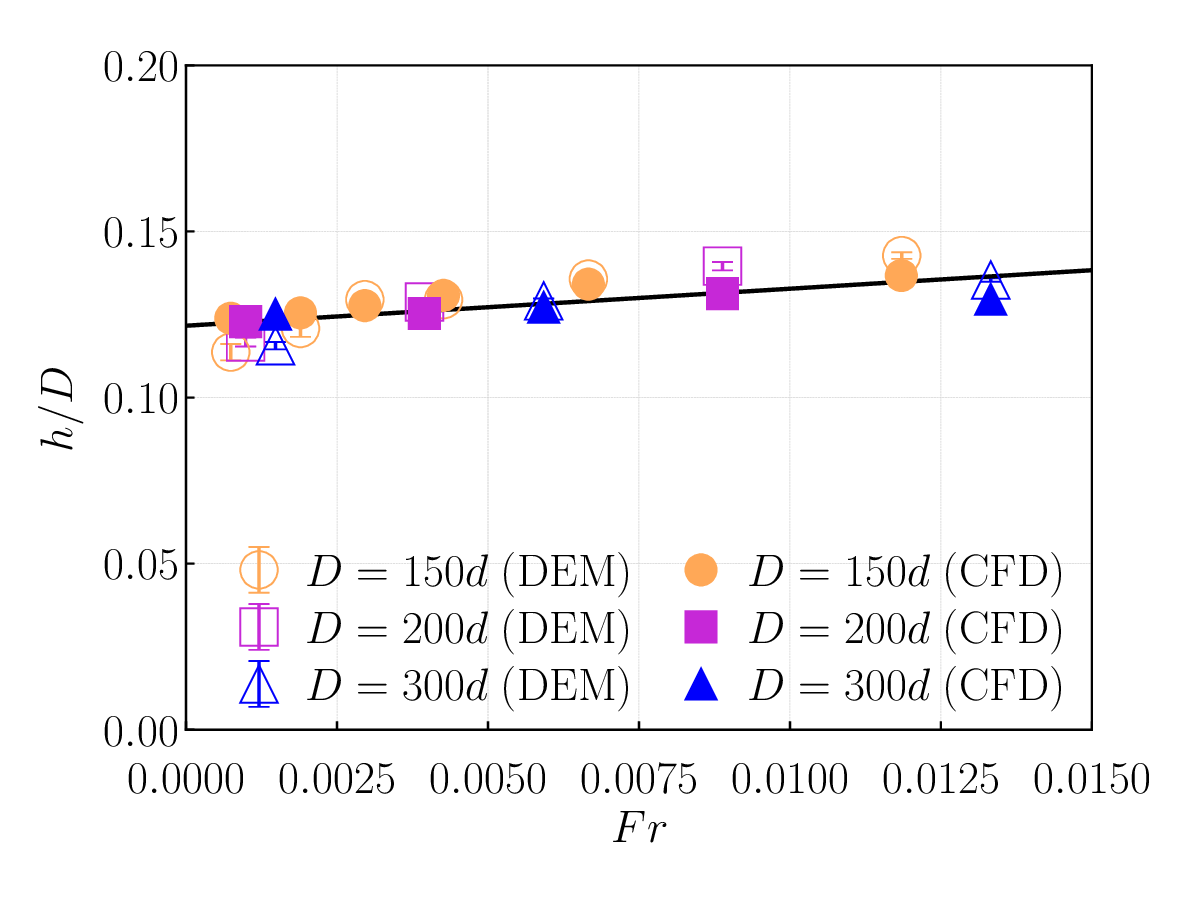}
    \caption{Normalized surface flow layer thickness $h/D$ against the Froude number $Fr$ for various values of $D$.
    The open and closed symbols represent the results of the DEM and CFD simulations, respectively.
    The solid line represents Eq. \eqref{eq:H_expand} with $H^{(0)} = 1.1 $ and $H^{(1)} = 1.2$, which are estimated by the least squares method.}
    \label{fig:Fr-h}
\end{figure}

Next, we plot $h/D$ against $Fr$ for various values of $d/D$ in Fig. \ref{fig:Fr-h}.
Closed and open symbols represent the results of the CFD and DEM simulations, respectively.
This figure confirms the validity of the scaling law, Eq. \eqref{eq:h_Fr}.
We have focused on the steady flow regime with low $Fr < 10^{-2}$.
In this regime, the scaling function $H(Fr)$ in Eq. \eqref{eq:h_Fr} can be expanded as
\begin{equation}
    H \left( Fr\right) = H^{(0)} + Fr H^{(1)} + O(Fr^2),
    \label{eq:H_expand}
\end{equation}
where $H^{(0)}$ and $H^{(1)}$ are constants.
As shown in Appendix \ref{appG}, $H^{(0)} > 0$ is required from flux conservation in the steady flow.
The solid line in Fig. \ref{fig:Fr-h} represents Eq. \eqref{eq:H_expand} with $H^{(0)} = 1.1$ and $H^{(1)} = 1.2$, estimated via the least squares method.
The agreement of the data with the solid line further supports the validity of the scaling law.
The scaling laws \eqref{eq:h_Fr} and \eqref{eq:H_expand} derived from the continuum model indicate that $h/D$ is nearly constant for $Fr \ll 1$, as shown in Fig. \ref{fig:Fr-h}.
Accordingly, the surface flow layer thickness $h$ is proportional to $D$ but independent of $\Omega$ in this regime.
This dependence is consistent with previous experimental results. \cite{Parker1997}

%----------------------------------------------%
% Discussion and conclusion                    %
%----------------------------------------------%
\section{Dicussion and Conclusion}
\label{sec:DiscussionAndConclusion}

In this paper, we aimed to understand how the surface flow layer thickness $h$ and the velocity profile of granular material in a two-dimensional rotating drum depend on the drum diameter $D$ and angular velocity $\Omega$.
To achieve this, we introduced a continuum model based on the $\mu(I)$-rheology and performed numerical simulations (CFD), complemented by discrete element method (DEM) simulations for validation.
This combined approach allowed us to quantitatively reproduce the velocity fields observed in the DEM simulations and to capture the overall flow behavior, including the rigid rotation in the static regime near the drum wall and the flow parallel to the free surface in the surface layer.
From dimensional analysis of the continuum equations, we analytically derived scaling laws for the velocity profile (Eq. \eqref{eq:u_Fr}) and the surface layer thickness (Eq. \eqref{eq:h_Fr}).
The numerical results confirm the validity of these scaling laws, demonstrating that the surface flow thickness $h$ is proportional to $D$ and nearly independent of $\Omega$ for sufficiently small Froude numbers.
Therefore, we conclude that the derived scaling laws are valid for 2D flows in half-filled rotating drums with large $D/d$ and low Froude numbers ($10^{-4} < Fr < 10^{-2}$), where the local and incompressible $\mu(I)$-rheology provides an appropriate continuum description.

The $\mu(I)$-rheology has been shown to be applicable to both two-dimensional \cite{DaCruiz2005} and three-dimensional granular systems \cite{Jop2006,Peyneau2008}.
When applying a continuum model based on the $\mu(I)$-rheology to granular flows in three-dimensional rotating drums, the scaling laws derived in this study, Eqs. \eqref{eq:u_Fr} and \eqref{eq:h_Fr}, are expected to hold provided that the flow is approximately uniform along the axial direction.
In such cases, the same dimensional arguments apply, although the scaling functions $U(z/D;Fr)$ and $H(Fr)$ may differ quantitatively due to changes in packing fraction and rheological parameters between 2D and 3D systems.

The scaling law for the surface flow layer thickness, Eq. \eqref{eq:h_Fr}, predicts $h \sim \mathrm{const.}$ in the limit $\Omega \to 0$, which is consistent with the experimental results reported in Ref.~\onlinecite{Parker1997}, although the estimated value of $H^{(0)}$ differs from that obtained in our two-dimensional simulations.
By contrast, several experimental studies using axially short rotating drums \cite{Khakhar1997,Felix2007,Orpe2007,Pignatel2012} have reported a power-law dependence $h \sim \Omega^\alpha$, based on phenomenological assumptions such as a constant shear rate.

Our preliminary three-dimensional simulations indicate that, for sufficiently long rotating cylinders, the flow remains nearly uniform along the axial direction except in narrow regions near the sidewalls.
In this regime, the scaling laws derived from the two-dimensional analysis are effectively satisfied.
For shorter cylinders, however, sidewall friction induces pronounced axial inhomogeneities, leading to deviations from the predicted scaling behavior.
These results suggest that sidewall effects \cite{Berzi2011} play a key role in the breakdown of the scaling laws and may account for the discrepancies between our predictions and experiments conducted in axially thin drums.
Further investigation is required to quantitatively characterize these effects and to extend the theoretical framework to fully three-dimensional geometries.

In this study, we considered a rotating drum half-filled with granular material.
Nevertheless, similar flow patterns as shown in Fig. \ref{fig:model} are generally observed for a range of fill fractions in previous studies.
We have not explicitly investigated the dependence of the velocity field on the fill fraction, and the derivation and confirmation of scaling laws for different fill fractions will be left for future work.

We have used the local $\mu(I)$-rheology with incompressible conditions in the continuum model.
Recent studies show that continuum models with the local, incompressible $\mu(I)$-rheology may exhibit ill-posed behavior, particularly in the low- and high-inertial-number ($I$) regimes \cite{Barker2015, Heyman2017}.
In our simulations, the high-$I$ regime does not occur, and for the low-$I$ region, a numerical cutoff is introduced for the effective viscosity $\eta_{\mathrm{g}}$ (see Appendix \ref{subsubsec:VOF}), which mitigates the ill-posedness.
Nevertheless, further investigation is needed to fully clarify the dependence on numerical parameters.

The $\mu(I)$-rheology has been extended to include non-local effects, which lead to deviations from a simple local constitutive law and generate exponential creep tails near the interface between the surface flow layer and the static flow regime \cite{Kamrin2012, Heyman2017, Kim2020}.
In the present study, such effects are neglected.
Analytical and numerical studies have shown that non-local effects become negligible for sufficiently large systems in granular flows between rough parallel plates \cite{Otsuki2025}.
Although an analogous analytical demonstration is not yet available for rotating drums, we expect that a similar argument applies in the limit of large drums with sufficiently large $D/d$.
Consistent with this expectation, the scaling laws derived from the local $\mu(I)$-rheology are confirmed in our simulations for two-dimensional rotating drums with $D/d \ge 150$.
In contrast, for smaller drums or three-dimensional systems with frictional side walls, non-local effects are expected to play a more significant role and may modify the scaling laws.
Such effects could contribute to the discrepancies between our predictions and the experimental results reported in Refs. \onlinecite{Khakhar1997, Felix2007, Orpe2007, Pignatel2012}.
Further studies are required to clarify the influence of non-local effects and sidewall friction on the velocity field in rotating drums.

Although we have obtained the scaling laws through non-dimensionalization, the explicit forms of the scaling functions have not yet been derived analytically.
Analytical studies on related systems provide useful guidance: for example, Haji-Sheikh and Elliot \cite{Haji-Sheikh1984} calculated the velocity field of an incompressible Newtonian fluid in a rotating drum, and Gray et al. \cite{Gray2001} derived the streamlines of granular flow using a simplified constitutive law.
Extending such analytical approaches to the present system could offer a deeper theoretical understanding of granular flows and help to explicitly determine the scaling functions.

In conclusion, the novelty of the present work lies not merely in reproducing velocity fields observed in experiments or DEM simulations, but in providing a theoretical framework that clarifies how the surface flow layer thickness $h$ and the velocity profile in rotating drums depend on the system parameters.
By applying dimensional analysis to the continuum equations based on the local, incompressible $\mu(I)$-rheology, we demonstrate that, in the limit of large $D/d$ and sufficiently small Froude numbers, the two-dimensional flow is governed primarily by the Froude number, leading to scaling laws in which the surface layer thickness is proportional to the drum diameter and nearly independent of the angular velocity.
This result helps to rationalize apparent discrepancies among previous experimental and numerical studies by suggesting that reported power-law dependencies on $\Omega$ arise from additional dimensionless parameters not captured in the idealized two-dimensional limit, such as sidewall friction, finite axial length, or non-local rheological effects.
In this sense, the present scaling analysis provides new physical insight into the origin and limitations of commonly observed scaling behaviors in granular flows within rotating drums and offers a systematic basis for interpreting deviations from the ideal scaling regime.

\begin{acknowledgments}
The authors thank S. Goto, Y. Motoori, D. Watanabe, and I. Procaccia for fruitful discussions.
We would like to thank Editage (www.editage.jp) for English language editing.
This study was supported by JSPS KAKENHI under Grant No. JP23K03248 and JP21H01006.
\end{acknowledgments}

\section*{Author declarations}
\subsection*{Conflict of Interest}

The authors have no conflicts to disclose.

\subsection*{Author Contributions}
\textbf{Hiroki Oba:} Conceptualization (equal); Data curation (lead); Formal Analysis (lead); Methodology (equal); Software (lead); visualization (lead); Writing -- original draft (lead). \textbf{Michio Otsuki} Conceptualization (equal);  Methodology (equal); Supervision (lead); Writing -- review \& editing (lead). 

\section*{Data Availability Statement}

The data that support the findings of this study are available from the corresponding author upon reasonable request.

\appendix

\section{Details of DEM simulations}
\label{App:DetailOfDEM}

In this Appendix, we present the details of our DEM simulations.
The dynamics of particle $i$ is given by the time evolution equations for the position $\vb*{r}_i = (X_i, Z_i)$ and angular velocity $\omega_i$ of particle $i$ as follows:
\begin{align}
    m_i \frac{d^2 \vb*{r}_i}{d t^2} &= \sum_{j \neq i} \vb*{F}_{ij} + m_i \vb*{g} , \\
    I^{\rm p}_{i} \frac{d \omega_i}{dt} &= \sum_{j \neq i} T_{ij} 
\end{align}
with $\vb*{g} = (0,-g)$.
Here, $m_i$ and $I^{\rm p}_i$ are the mass and moment of inertia of particle $i$, respectively.
The contact force and torque between particle $i$ and particle $j$ are denoted by $\vb*{F}_{ij}$ and $T_{ij}$, respectively.

The contact force $\vb*{F}_{ij}$ consists of the normal component $\vb*{F}^{(\mathrm{n})}_{ij}$ and the tangential component $\vb*{F}^{(\mathrm{t})}_{ij}$ as follows:
\begin{equation}
    \vb*{F}_{ij} = \vb*{F}^{(\mathrm{n})}_{ij} + \vb*{F}^{(\mathrm{t})}_{ij} .
    \label{eq:DecompositionOfForce}
\end{equation}
The normal contact force $\vb*{F}^{(\mathrm{n})}_{ij}$ is given by 
\begin{equation}
    \vb*{F}^{(\mathrm{n})}_{ij} = \left( - k^{(\mathrm{n})} \vb*{\delta}^{(\mathrm{n})}_{ij}
    - \eta^{(\mathrm{n})}_{ij} \vb*{v}^{(\mathrm{n})}_{ij} \right)
    \Theta \left( \delta_{ij} \right) ,
    \label{eq:NormalContactForce}
\end{equation}
where $k^{(\mathrm{n})}$ and $\eta^{(\mathrm{n})}_{ij}$ are the normal elastic and the normal viscous constants, respectively;
here, $\Theta(x)$ is the Heaviside step function, defined by $\Theta(x) = 1$ for $x \geq 0$ and $\Theta(x) = 0$ for $x < 0$. 
The normal displacement $\vb*{\delta}^{(\mathrm{n})}_{ij}$, the normal relative velocity $\vb*{v}^{(\mathrm{n})}_{ij}$, and the normal unit vector $\vb*{n}_{ij}$ in Eq. \eqref{eq:NormalContactForce} are respectively given by
\begin{gather}
    \vb*{\delta}^{(\mathrm{n})}_{ij} = \delta_{ij} \vb*{n}_{ij} , 
    \label{eq:NormalDisplacement} \\
    \vb*{v}^{(\mathrm{n})}_{ij} = \left\{(\vb*{v}_i - \vb*{v}_j) \cdot \vb*{n}_{ij}\right\} \vb*{n}_{ij} , 
    \label{eq:NormalRelativeVelocity} \\
    \vb*{n}_{ij} = \frac{\vb*{r}_j - \vb*{r}_i}{\left|\vb*{r}_j - \vb*{r}_i\right|} ,
    \label{eq:NormalVector}
\end{gather}
where $\vb*{v}_i$ is the velocity of particle $i$, and $\delta_{ij}$ is the overlap between particles $i$ and $j$ given by
\begin{equation}
    \delta_{ij} = \frac{d_i + d_j}{2} - \left| \vb*{r}_j - \vb*{r}_i \right|
    \label{eq:ConditionOfContact}
\end{equation}
with the diameter $d_i$ of particle $i$.

The tangential contact force $\vb*{F}^{(\mathrm{t})}_{ij}$ obeys the Coulomb friction law as follows:
\begin{equation}
    \vb*{F}^{(\mathrm{t})}_{ij} = 
    \mathrm{min} \left( \left| \vb*{F}^{(\mathrm{t,st})}_{ij} \right|, \;
    \mu_{\mathrm{p}} \left| \vb*{F}^{(\mathrm{n})}_{ij} \right| \right)
    \frac{\vb*{F}^{(\mathrm{t,st})}_{ij}}{\left| \vb*{F}^{(\mathrm{t,st})}_{ij} \right|}
    \Theta \left( \delta_{ij} \right) 
    \label{eq:TangentialContactForce}
\end{equation}
with the particle friction coefficient $\mu_{\mathrm{p}}$.
Here, $\mathrm{min} (a,b)$ in Eq. \eqref{eq:TangentialContactForce} selects the smaller of $a$ and $b$, and
$\vb*{F}^{(\mathrm{t,st})}_{ij}$ is given by
\begin{equation}
    \vb*{F}^{(\mathrm{t,st})}_{ij} = 
    - k^{(\mathrm{t})} \delta^{(\mathrm{t})}_{ij} \vb*{t}_{ij}
    - \eta^{(\mathrm{t})}_{ij} \vb*{v}^{(\mathrm{t})}_{ij}
    \label{eq:TangentialContactForce_stick}
\end{equation}
with the tangential spring constant $k^{(\mathrm{t})}$ and the tangential viscous constant $\eta^{(\mathrm{t})}_{ij}$.
The tangential displacement ${\delta}^{(\mathrm{t})}_{ij}$, the tangential relative velocity $\vb*{v}^{(\mathrm{t})}_{ij}$, and the tangential unit vector $\vb*{t}_{ij}$ are given by
\begin{gather}
    \delta^{(\mathrm{t})}_{ij} = \int_{\mathrm{stick}} dt \ \vb*{v}^{(\mathrm{t})}_{ij} \cdot \vb*{t}_{ij} , 
    \label{eq:TangentialDisplacement} \\
    \vb*{v}^{(\mathrm{t})}_{ij} = \left\{ \left( \vb*{v}_i - \vb*{v}_j \right) \cdot \vb*{t}_{ij} + \frac{1}{2} \left( d_i \omega_i + d_j \omega_j \right) \right\} \vb*{t}_{ij} , 
    \label{eq:TangentialRelativeVelocity} \\
    \vb*{t}_{ij} = \left( - n_{ij,Z}, \; n_{ij,X} \right).
    \label{eq:TangentialVector}
\end{gather}
The subscript ``$\mathrm{stick}$'' in Eq. \eqref{eq:TangentialDisplacement} indicates that the integral is performed only when $\left| \vb*{F}^{(\mathrm{t,st})}_{ij} \right| <  
    \mu_{\mathrm{p}} \left| \vb*{F}^{(\mathrm{n})}_{ij} \right| $.
The torque $T_{ij}$ is given by
\begin{equation}
    T_{ij} = \frac{d_i}{2} \vb*{F}^{(\mathrm{t})}_{ij} \cdot \vb*{t}_{ij} .
    \label{eq:Torque}
\end{equation}

The diameters of the particles are uniformly distributed from $d$ to $1.2d$.
The mass and moment of inertia of particle $i$ are given by $m_i = \frac{1}{4} \pi \rho_s d_i^2$ and $I^p_i=\frac{1}{8} m_i {d_i}^2$, respectively.
We use $k^{(\mathrm{n})}=2.0 \times 10^5 mg/d$, $k^{(\mathrm{t})} / k^{(\mathrm{n})} = 2 / 7$, and $\mu_p=0.50$.
The normal viscous constant between particles $i$ and $j$ is given by
\begin{gather}
    \eta^{(\mathrm{n})}_{ij} = \eta^{(\mathrm{t})}_{ij} = -2 \ln{e} \sqrt{\frac{m^*_{ij} k^{(\mathrm{n})}}{\pi^2 + (\ln{e})^2}}
    \label{eq:DefinitionOfViscoelasticConstant} 
\end{gather}
with the coefficient of restitution $e=0.92$ and
\begin{gather}
    m^*_{ij} = \frac{m_i m_j}{m_i + m_j}.
    \label{eq:DefinitionOfReducedMass}
\end{gather}
The time evolution equations are numerically solved using the leapfrog algorithm with the time step $\Delta t = 1.0 \times 10^{-4} \sqrt{d/g}$.

The rough wall surface of the drum is modeled by placing $N_{\rm w}$ particles of diameter $d$ equally spaced along a circle of diameter $D$ in order to prevent slip at the boundary.
These wall particles are rigidly rotated with the angular velocity $\Omega$.
We set $N_{\rm w}=475$, $632$, and $946$ for drums with $D=150d$, $300d$, and $450d$, respectively.
The drum is half-filled with $N$ particles, where $N=7700$, $30600$, and $69200$ for $D=150d$, $300d$, and $450d$, respectively.
To generate the initial packing, the $N$ particles are first placed randomly inside the drum and allowed to settle under gravity without rotation.
Subsequently, the drum rotation with angular velocity $\Omega$ is started and the system is evolved until a steady flow state is reached.
In the DEM simulations, the packing fraction is not explicitly controlled but is instead determined self-consistently by the particle dynamics under gravity.

\section{Velocity field in DEM simulation}
\label{appA}
The velocity distribution in Fig. \ref{fig:velocity_vmap}(a) is calculated using the following procedure in the DEM simulation.
First, the system is divided into small square regions with a length of $4.8d$ in the plane of the Cartesian coordinates $\left( X, Z \right)$.
Then, the average particle velocity in each region is plotted as the velocity field.
The velocity field in the plane of the Cartesian coordinates $(x,z)$ shown in Fig. \ref{fig:velocity_center} is numerically obtained as the average particle velocity in the square regions with a length of $1.2d$.

\section{Details of CFD simulation}
\label{App:CFD}
In this Appendix, we present the details of our CFD simulation.
In the CFD simulation, we treat the granular bed in the rotating drum as an incompressible non-Newtonian fluid.
The continuum equations for the granular material given by Eqs. \eqref{eq:Mass}--\eqref{eq:Viscosity} are numerically solved using the finite difference method with the time step $\Delta t$ and grid width $\Delta x$.
The free space above the granular bed is modeled as a gas phase with very low density and viscosity, allowing the free surface of the granular material to be treated.
The time evolution of the two-phase flow with the granular and gas phases is numerically solved by the volume of fluid (VoF) method. \cite{Hirt1981,Gueyffier1999,Li2000}

\subsection{Governing equations}
\label{subsubsec:GoverningEquation}
The entire domain $\mathrm{V}$ of the system is expressed as $\mathrm{V} = \mathrm{V_f} \cup \mathrm{V_s}$ with the fluid domain $\mathrm{V_f}$ including the granular and gas phases and the solid domain $\mathrm{V_s}$ representing the drum wall.
The mass conservation equation is given by
\begin{equation}
    \nabla \cdot \vb*{v} = 0 \; \; \; \; \; \mathrm{in} \; \; \; \vb*{r} \in \mathrm{V} ,
    \label{eq:MassConservation}
\end{equation}
while the momentum conservation equation in the fluid domain is given by
\begin{equation}
    \frac{\partial \vb*{v}}{\partial t} = - \nabla \cdot \left( \vb*{v} \vb*{v} \right) - \frac{1}{\rho} \nabla p + \frac{1}{\rho} \nabla \cdot \vb*{\tau} + \vb*{g} \; \; \; \; \; \mathrm{in} \; \; \; \vb*{r} \in \mathrm{V_f} ,
    \label{eq:MomentumConservation}
\end{equation}
where $\rho$, $\vb*{v}$, and $p$ are the density, velocity, and pressure at $\vb*{r} = (X,Z)$ and time $t$, respectively.
Here, $\vb*{\tau}$ is given by
\begin{equation}
    \vb*{\tau} = \eta \vb*{\dot{\gamma}}
    \label{eq:DefinitionOfShearStress}
\end{equation}
with the viscosity $\eta$ and shear rate tensor $\vb*{\dot{\gamma}} = \nabla \vb*{v} + (\nabla \vb*{v})^T$.
Notably, $\eta$ depends on the position in the fluid domain.
The velocity vector in the solid domain is given by
\begin{equation}
    \vb*{v} = \vb*{v}_{\mathrm{wall}} \; \; \; \; \; \mathrm{in} \; \; \; \vb*{r} \in \mathrm{V_s} ,
    \label{eq:MomentumConservation_Wall}
\end{equation}
with $\vb*{v}_{\mathrm{wall}}$ given by Eq. \eqref{eq:WallVelocity}.

\subsection{Volume of fluid method}
\label{subsubsec:VOF}

In the VoF method, we introduce the phase volume fraction $\alpha(\vb*{r},t)$ of the fluid phase.
Here, $\alpha=1$ and $\alpha=0$ correspond to the areas occupied only by the granular and gas phases, respectively.
In the interface between the granular and gas phases, $\alpha$ satisfies $0<\alpha<1$.
We consider the position with $\alpha=0.5$ as the free surface of the granular bed.
The phase volume fraction $\alpha(\vb*{r},t)$ obeys the advection equation
\begin{equation}
    \frac{\partial \alpha}{\partial t} + \nabla \cdot \left( \alpha \vb*{v} \right) - \alpha \nabla \cdot \vb*{v} = 0 .
    \label{eq:VoF-ConvectionEquation}
\end{equation}
This equation is solved using the WLIC/THINC (weighed line interface calculation/tangent of hyperbola for interface capturing) method. \cite{Yokoi2007}
The density $\rho$ and viscosity $\eta$ are calculated as follows:
\begin{align}
    \rho &= \rho_{\mathrm{g}} \alpha + \rho_{\mathrm{air}} \left( 1-\alpha \right) ,
    \label{eq:RhoDefinition} \\
    \eta &= \eta_{\mathrm{g}} \left( \dot{\gamma}, p \right) \alpha + \eta_{\mathrm{air}} \left( 1-\alpha \right) .
    \label{eq:EtaDefinition}
\end{align}
Here, $\rho_{\mathrm{air}}$ and $\eta_{\mathrm{air}}$ are constants representing the density and viscosity in the gas phase, respectively. 
The density $\rho_{\mathrm{g}}$ in the granular phase is treated as a constant.
The viscosity $\eta_{\mathrm{g}}$ in the granular phase is given by Eq. \eqref{eq:Viscosity} in the $\mu(I)$-rheology. \cite{Jop2006}
We adopt the functional form of $\mu(I)$ as Eq. \eqref{eq:mu_I_Rheology_Function}.
The inertial number $I$ is given by $I = \dot{\gamma} / \sqrt{p/(\rho_s d^2)}$ with
\begin{equation}
    \dot{\gamma} = \left( \frac{1}{2} \vb*{\dot{\gamma}} : \vb*{\dot{\gamma}} \right)^{\frac{1}{2}} ,
    \label{eq:DefinitionOfNorm}
\end{equation}
which corresponds to the second invariant of $\vb*{\dot{\gamma}}$.

The viscosity in Eq. \eqref{eq:Viscosity} diverges in the limit of $\dot{\gamma} \to 0$.
To prevent the divergence in the CFD simulation, we regularize Eq. \eqref{eq:Viscosity} by approximating it with a cutoff $\eta_{\mathrm{g,max}}$ \cite{Arseni2020} as follows:
\begin{equation}
    \eta_{\mathrm{g}} = \mathrm{min} \left[ \frac{\mu(I) p}{\dot{\gamma}}, \: \eta_{\mathrm{g,max}} \right] .
    \label{eq:EtaOfGranular}
\end{equation}
The rigid rotation in the static flow regime is effectively realized using a sufficiently large value for $\eta_{\mathrm{g,max}}$.

\subsection{Simplified marker and cell method}
\label{subsubsec:SMAC}
In this study, we use the simplified marker and cell (SMAC) method \cite{Amsden1970} to solve the time evolution equations in the fluid domain.
We denote an arbitrary physical quantity $\vb{a}$ at the time $t=n\Delta t$ by $\vb{a}^n$ with the time step $\Delta t$ and an integer $n$.
In the SMAC method, $\vb{a}^{(n+1)}$ at the time $t=(n+1)\Delta t$ is numerically obtained from $\vb{a}^{(n)}$ in two steps.

%The SMAC method evolves the system in time by repeating two steps: a prediction step in which the momentum conservation equation is explicitly solved to calculate the predicted velocity, and a modification step in which it is corrected to satisfy the continuity equation.

First, for the prediction step, the predicted velocity $\vb*{v}^{*}$ in $\vb*{r} \in \mathrm{V_f}$ is determined by
\begin{equation}
    \frac{\vb*{v}^{*} - \vb*{v}^{n}}{\Delta t} = \frac{3 \vb*{A}^{n} - \vb*{A}^{n-1}}{2} - \frac{1}{\rho^{n}} \nabla S^{n} + \frac{\vb*{B}^{*} + \vb*{B}^{n}}{2} + \vb*{g} ,
    \label{eq:SMAC-1_f}
\end{equation}
Here, we have used the second-order Adams--Bashforth method for the advection term 
\begin{align}
    &\vb*{A} = - \nabla \cdot \left( \vb*{vv} \right)
    \label{eq:ConvectionTerm}, 
\end{align}
while the second-order Crank--Nicolson method is used for the viscous term
\begin{align}
    &\vb*{B} = \frac{1}{\rho} \nabla \cdot \left[ \eta \left( \nabla \vb*{v} + (\nabla \vb*{v})^T \right) \right]
    \label{eq:ViscosityTerm} .
\end{align}
In the SMAC method, a scholar potential $S$ is introduced in Eq. \eqref{eq:SMAC-1_f} to satisfy the mass conservation.
In Eq. \eqref{eq:SMAC-1_f}, $\vb*{B}^{*}$ is given by  
\begin{equation}
    \vb*{B}^{*} = \nu_0 \nabla^{2} \vb*{v}^{*} + \frac{1}{\rho^{n}} \nabla \cdot \left[ \eta^{n} \left( \nabla \vb*{v}^{n} + (\nabla {\vb*{v}^{n}})^T \right) \right] - \nu_0 \nabla^{2} \vb*{v}^{n},
    \label{eq:Dodd-et-al2014}
\end{equation}
where the first term is treated implicitly, and the other terms are treated explicitly. \cite{Dodd2014}
Here, $\nu_0$ is the constant component of kinematic viscosity given by
\begin{equation}
    \nu_0 = \frac{1}{2} \left( \frac{\eta_{\mathrm{air}}}{\rho_{\mathrm{air}}} + \frac{\eta_{\mathrm{g,max}}}{\rho_{\mathrm{g}}} \right) .
    \label{eq:AverageViscosity}
\end{equation}
For $\vb*{r} \in \mathrm{V_s}$, the predicted velocity $\vb{v}^{*}$ is given by
\begin{equation}
    \vb*{v}^{*} = \vb*{v}_{\mathrm{wall}} .
    \label{eq:SMAC-Wall_step1}
\end{equation}

Next, for the projection step, $\vb*{v}^{\mathrm{n+1}}$ in $\vb*{r} \in \mathrm{V_f}$ is obtained as follows:
\begin{equation}
    \vb*{v}^{\mathrm{n+1}} = \vb*{v}^{*} - \frac{\Delta t}{\rho^{n}} \nabla \sigma .
    \label{eq:SMAC-2_f}
\end{equation}
Here, $\sigma$ is given by the solution of the Poisson equation
\begin{equation}
    \nabla \cdot \left( \frac{1}{\rho^{n}} \nabla \sigma \right) = \frac{1}{\Delta t} \nabla \cdot \vb*{v}^{*} 
    \label{eq:SMAC-Poisson}
\end{equation}
with the boundary condition $\nabla \sigma = \vb*{0}$ at the wall surface, which is derived by taking the divergence of  Eq. \eqref{eq:SMAC-2_f} with the mass conservation $\nabla \cdot \vb*{v}^{n+1} = 0$.
The scholar potential $S^{n+1}$ is updated as follows:
\begin{equation}
    S^{n+1} = S^{n} + \sigma - \frac{1}{2} \Delta t \nu_0 \nabla^2 \sigma .
    \label{eq:TimeEvolutionOfPressure}
\end{equation}
In $\vb*{r} \in \mathrm{V_s}$, $\vb*{v}^{\mathrm{n+1}}$ is given by
\begin{equation}
    \vb*{v}^{\mathrm{n+1}} = \vb*{v}^{*} .
    \label{eq:SMAC-Wall_step2}
\end{equation}

%Since the scholar potential $S^n$ in the SMAC method does not represent the physically correct pressure \citep{Amsden1970}, 
According to Ref. \onlinecite{Amsden1970},
we calculate the pressure $p^n$ used in Eq. \eqref{eq:EtaOfGranular} as the solution of 
\begin{equation}
    \nabla \cdot \left\{ \frac{1}{\rho^n} \nabla p^n \right\} = \nabla \cdot \left( - \nabla \cdot \left( \vb*{v}^n \vb*{v}^n \right) + \frac{1}{\rho^n} \nabla \cdot \vb*{\tau}^n + \vb*{g} \right) 
    \label{eq:PressureEquation}
\end{equation}
with the boundary conditions $\nabla p = \alpha \rho \vb*{g}$ at the wall surface, which is derived from the divergence of the momentum conservation Eq. \eqref{eq:MomentumConservation}.

\subsection{Boundary data immersion method}
\label{subsubsec:BDI}
We treat the flow driven by the rotating drum using the boundary data immersion (BDI) method. \cite{Weymouth2011} The solid domain $\mathrm{V_s}$ is defined as the circular shell with the inner and outer diameters $D$ and $D + 16 \Delta x$, respectively.
In this method, we introduce a single meta-equation to solve the time evolution equations in the fluid and solid domains simultaneously.
Here, we define the kernel function
\begin{equation}
    \chi (d_k) =
    \begin{cases}
        1 \; \; &\mathrm{for} \; \; d_k \leq - \epsilon \\
        \frac{1}{2} \left[ 1 - \sin \left( \frac{\pi d_k}{2 \epsilon} \right) \right] \; \; &\mathrm{for} \; \; \left| d_k \right| < \epsilon \\
        0 \; \; & d_k \geq  \epsilon 
    \end{cases}
    \label{eq:KernelFunction_BDIM}
\end{equation}
to continuously connect the fluid and solid domains.
Here, the shortest distance from the wall surface at the position $\vb*{r}$ is denoted by $d_k(\vb*{r})$, and $\epsilon$ is a parameter characterizing the thickness of the interface between the fluid and solid domains.
From Eq. \eqref{eq:KernelFunction_BDIM}, the kernel function $\chi$ is $1$ in the fluid domain and $0$ in the solid domain.

Using the BDI method with Eqs. \eqref{eq:SMAC-1_f} and \eqref{eq:SMAC-Wall_step1}, the meta-equation for the predicted velocity $\vb*{v}^{*}$ is given by
\begin{equation}
    \vb*{v}^{*} = \chi \left[ \left( 1 - \frac{\nu_0 \Delta t}{2} \nabla^2 \right)^{-1} \vb*{\mathrm{RHS}}^{n} \right] + \left( 1 - \chi \right) \vb*{v}_{\mathrm{wall}} ,
    \label{eq:SMAC-1_BDIM}
\end{equation}
where
\begin{align}
    \vb*{\mathrm{RHS}}^{n} = \vb*{v}^{n} + \Delta t \left[ \frac{3 \vb*{A}^{n} - \vb*{A}^{n-1}}{2} - \frac{1}{\rho^{n}} \nabla S^{n} + \vb*{B}^{n} - \frac{\nu_0}{2} \nabla^{2} \vb*{v}^{n} + \vb*{g} \right] .
    \label{eq:SMAC-1_2_RHS}
\end{align}
For the projection step, the meta-equation for $\vb*{v}^{\mathrm{n+1}}$ is derived as follows:
\begin{equation}
    \vb*{v}^{\mathrm{n+1}} = \vb*{v}^{*} - \frac{\Delta t}{\rho^{n}} \chi \nabla \sigma
    \label{eq:SMAC-2_BDIM}
\end{equation}
from Eqs. \eqref{eq:SMAC-2_f} and \eqref{eq:SMAC-Wall_step2}.
The Poisson equation for $\sigma$ is obtained as follows:
\begin{equation}
    \nabla \cdot \left( \frac{\chi}{\rho^{n}} \nabla \sigma \right) = \frac{1}{\Delta t} \nabla \cdot \vb*{v}^{*} ,
    \label{eq:SMAC-Poisson_Meta}
\end{equation}
from the divergence of Eq. \eqref{eq:SMAC-2_BDIM}.
The Poisson equation for $p^n$ is given by
\begin{eqnarray}
  \nabla \cdot \left\{ \frac{\chi}{\rho^n} \nabla p^n \right\} & = & \nabla \cdot \left\{ \chi \left( - \nabla \cdot \left( \vb*{v}^n \vb*{v}^n \right) + \frac{1}{\rho^n} \nabla \cdot \vb*{\tau}^n + \vb*{g} \right)\right\} \nonumber \\
  & & + \nabla \cdot \left\{ \left( 1 - \chi \right) \frac{\vb*{v}_{\mathrm{wall}}}{\Delta t} \right \},
    \label{eq:PressureEquation_Meta}
\end{eqnarray}
which is obtained from the divergence of
\begin{eqnarray}
  \vb*{v}^{n+1} & = & \chi \left[ \vb*{v}^n + \Delta t \left\{ - \nabla \left( \vb*{v}^n \vb*{v}^n \right) - \frac{1}{\rho^n} \nabla p^n + \frac{1}{\rho} \nabla \cdot \vb*{\tau}^n + \vb*{g} \right\} \right] \nonumber \\
  & & + \left( 1 - \chi \right) \vb*{v}_{\mathrm{wall}} 
\end{eqnarray}
with Eqs. \eqref{eq:MomentumConservation} and \eqref{eq:MomentumConservation_Wall}.
We solve Eqs. \eqref{eq:SMAC-Poisson_Meta} and \eqref{eq:PressureEquation_Meta} using the Bi-CGSTAB method. \cite{Vorst1992}

Note that near the interface between the granular and gas phases, the solution of Eq. \eqref{eq:PressureEquation_Meta} may result in pressure below $0$, which leads to unphysical behavior when calculating the viscosity of the granular phase $\eta_{\mathrm{g}}$ in Eq. \eqref{eq:EtaOfGranular}. \cite{Chauchat2014,Lin2020}
To prevent this, a sufficiently small positive constant $p_{\mathrm{min}}$ is set as the minimum value of the pressure.

\subsection{Numerical conditions}
\label{subsubsec:Setup_CFD}
We estimate $\mu_{s}$, $\mu_{2}$, and $I_{0}$ in Eq. \eqref{eq:mu_I_Rheology_Function} as $\mu_{s} = 0.246$, $\mu_{2} = 0.401$, and $I_{0} = 0.133$ from the DEM simulation under uniform shear with the parameters in Sec. \ref{subsec:DEM}. 
The details of the estimation are shown in Appendix \ref{appB}.
We set the granular density as $\rho_g = \rho_s \phi$ with $\rho_s = m / (\pi d^2 / 4)$.
The packing fraction $\phi$ of the particle is set to $0.825$ based on its value in the static flow regime of the rotating drum in the DEM simulation (Appendix \ref{appD}).
We use  $\eta_{\mathrm{g,max}} = 8000 m\sqrt{g/d}$, $\rho_{\mathrm{air}} = 10^{-3} \rho_g$, $\eta_{\mathrm{air}} = 10^{-6} \eta_{\mathrm{g,max}}$, and $p_{\mathrm{min}} = 10^{-5} mg/d$.
We have checked that the velocity profile is quantitatively the same even if we use larger $\eta_{\mathrm{g,max}}$ and lower $\rho_{\mathrm{air}}$, $\eta_{\mathrm{air}}$, and $p_{\mathrm{min}}$.
We set $\Delta x = D/75$, $\Delta t = 1.0 \times 10^{-4} \sqrt{d/g}$, and $\epsilon = 2 \Delta x$.
The maximum error for the Bi-CGSTAB method is set to $e_{\rm max}=10^{-5}$.

In this simulation, the free surface of the granular phase is initially set to be aligned with the $X$ axis, i.e., $\alpha = 1$ for $Z<0$ and $\alpha=0$ for $Z\ge 0$.
The initial velocity is $0$ in the entire system.
From this initial state, we generate a steady flow by rotating the drum counterclockwise with the angular velocity $\Omega$.

\section{Determination of parameters for $\mu(I)$-rheology}
\label{appB}
\begin{figure}
    \centering
    \includegraphics[width=1.0\hsize]{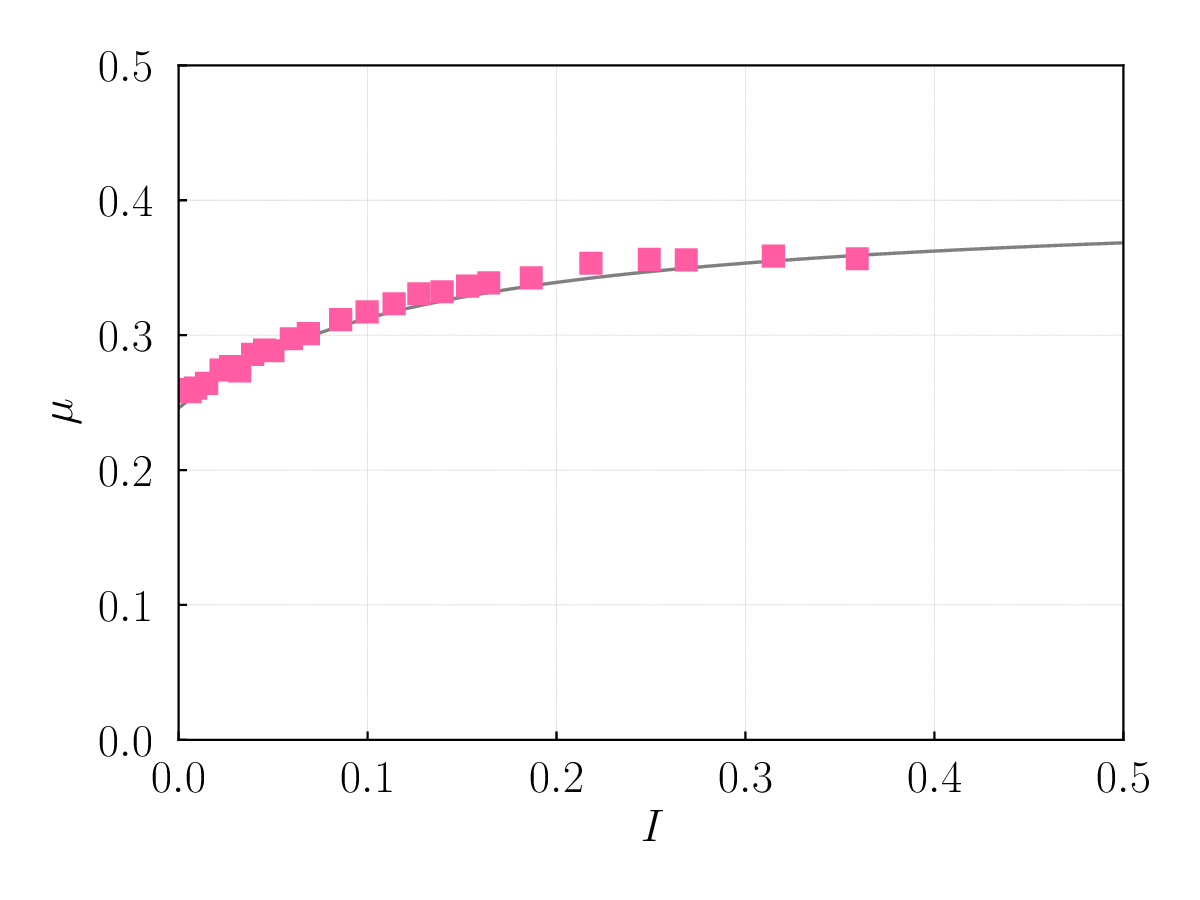}
    \caption{Bulk friction $\mu$ against the inertial number $I$ obtained from the DEM simulation under uniform shear.
    The solid line represents Eq. \eqref{eq:mu_I_Rheology_Function} with $\mu_s = 0.246$, $\mu_2 = 0.401$, and $I_0 = 0.133$.
    }
    \label{fig:DEM_LE_mu_I}
\end{figure}

To estimate the parameters in Eq. \eqref{eq:mu_I_Rheology_Function}, we have performed the DEM simulation for the granular materials shown in Sec. \ref{subsec:DEM} under uniform shear with the shear rate $\dot \gamma$ and packing fraction $\phi$ using the Lees-Edwards boundary condition. \cite{Lees1972}
We use $\phi=0.825$, which is the value in the static flow regime of the rotating drum.
In this simulation, the shear stress $\tau$ and pressure $p$ are measured, and we plot the bulk friction $\mu=\tau/p$ against the inertial number $I$ in Fig. \ref{fig:DEM_LE_mu_I}.
This figure shows that the $I$-dependence of $\mu$ is consistent with the theoretical expression, Eq. \eqref{eq:mu_I_Rheology_Function}, for the $\mu(I)$-rheology represented by the solid line.
We have determined the parameters as $\mu_s = 0.246$, $\mu_2 = 0.401$, and $I_0 = 0.133$ by fitting the numerical data with Eq. \eqref{eq:mu_I_Rheology_Function}.

\section{Packing fraction in the rotating drum}
\label{appD}
\begin{figure}
    \centering
    \includegraphics[width=1.0\hsize]{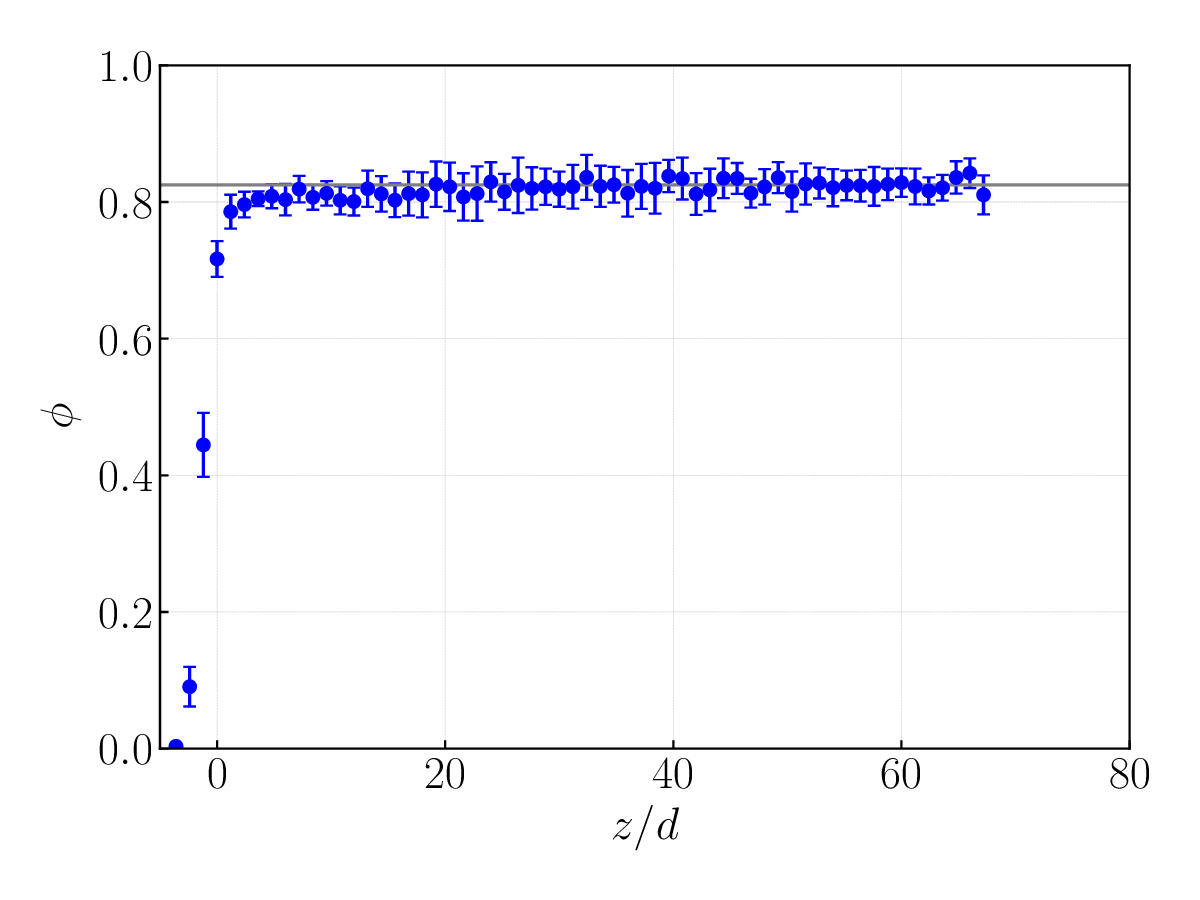}
    \caption{Packing fraction $\phi$ against $z$ at $x=0$ obtained from the DEM simulation with $D=150d$ and $\Omega = 6.28 \times 10^{-3} \sqrt{\rho_g g/d}$.
    The solid line represents the average value of $\phi$ in the solid region.
    }
    \label{fig:DEM_VF}
\end{figure}

Figure \ref{fig:DEM_VF} presents the dependence of the volume fraction $\phi$ on $z$ at $x=0$ obtained from the DEM simulation with  $D=150d$ and $\Omega = 6.28 \times 10^{-3} \sqrt{\rho_g g/d}$.
The packing fraction $\phi$ is nearly constant except for the region near $z \approx 0$.
The packing fraction in the static flow regime is estimated to be $0.825$, which is indicated by the straight line in Fig. \ref{fig:DEM_VF}, though the packing fraction in the surface flow layer is slightly small due to the dilatancy of granular materials under shear. \cite{Renouf2005}
Therefore, in the CFD simulation, the granular material is treated as an incompressible fluid.
In the DEM simulation, $\phi = 0.7$ is defined as the free surface position.

\section{Position of free surface}
\label{appE}
\begin{figure}
    \centering
    \includegraphics[width=1.0\hsize]{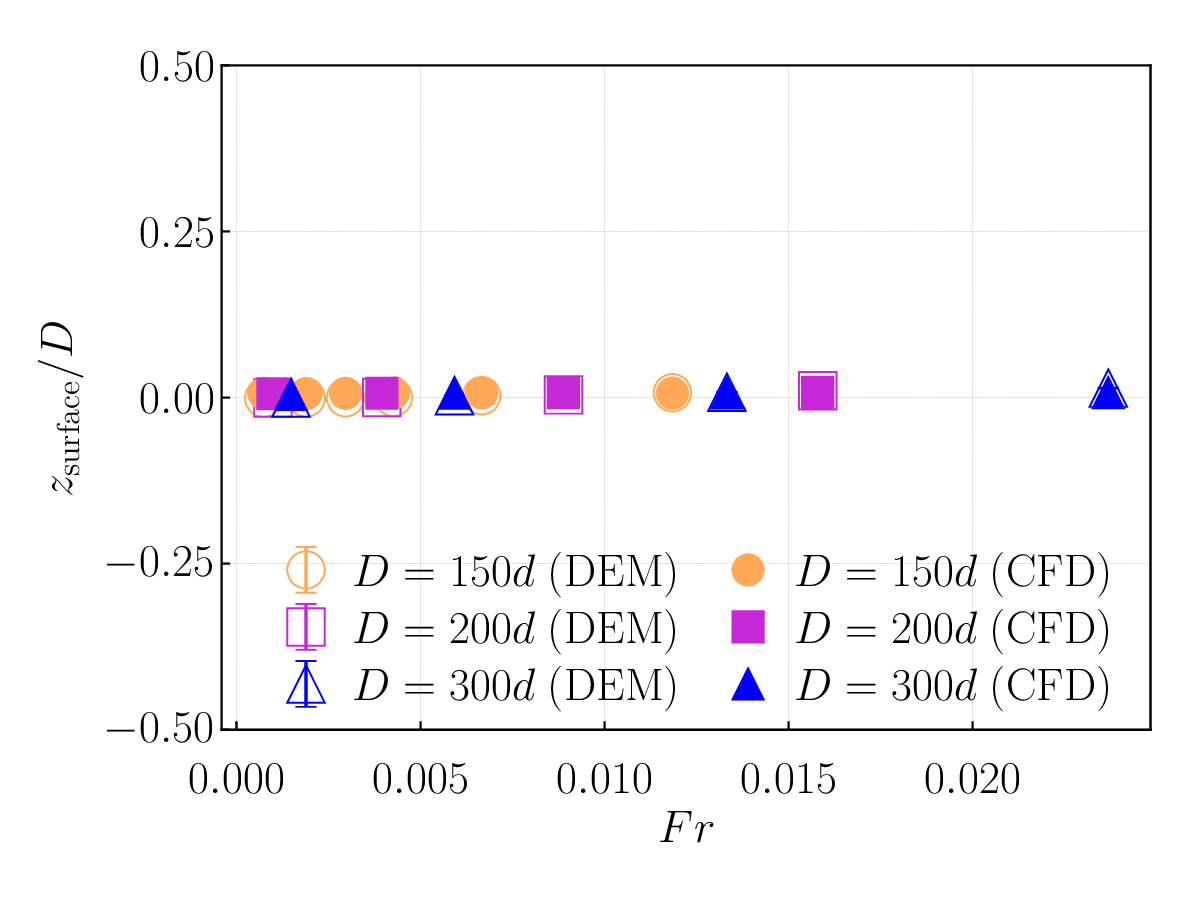}
    \caption{Normalized free surface height $z_{\mathrm{surface}}/D$ at $x=0$ against the Froude number $Fr$.
    The open and closed symbols represent the results of the DEM and CFD simulations, respectively.
    }
    \label{fig:free-surface}
\end{figure}

Figure \ref{fig:free-surface} illustrates the position of the free surface $z_{\mathrm{surface}}$ against the Froude number $Fr$ at $x=0$ for each $D$.
The open and closed symbols represent the results of the DEM and CFD simulations, respectively.
The normalized free surface position $z_{\mathrm{surface}}/D$ is close to $0$ for all of the Froude number $Fr$.
Therefore, we treat $z_{\mathrm{surface}} = 0$ in this study.

\section{Boundary between surface flow layer and static flow regime}
\label{appF}
In this study, the boundary between the surface flow layer and the static flow regime is defined as the position for $u(z)=0$. However, in Ref. \onlinecite{Ding2002}, the boundary is determined from the difference between $u(z)$ and the rigid rotation velocity $-\Omega z$.
For example, we may regard the surface flow layer as the region where the velocity in the granular bed is sufficiently larger than that of the rigid rotation as
\begin{equation}
    \frac{u(z) - \left( -\Omega z \right)}{\Omega D} \geq \alpha_{\mathrm{th}} 
    \label{eq:Boundary_Another}
\end{equation}
with a threshold $\alpha_{\mathrm{th}}$.
The surface flow layer thickness $h^{\prime}$ for this definition satisfies
\begin{equation}
    \frac{u(z = h^{\prime}) - \left( -\Omega h^{\prime} \right)}{\Omega D} = \alpha_{\mathrm{th}} .
    \label{eq:h_Another}
\end{equation}
If the velocity $u(z)$ obeys the scaling law, Eq. \eqref{eq:u_Fr}, this equation gives
\begin{equation}
    U \left( \frac{h^{\prime}}{D} ; Fr \right) - \frac{h^{\prime}}{D} = \alpha_{\mathrm{th}} .
\end{equation}
Solving this equation for $h^{\prime}/D$ yields the scaling law for $h^{\prime}$ as
\begin{equation}
    \frac{h^{\prime}}{D} = H^{\prime} \left( Fr \right) 
\end{equation}
for a given $\alpha_{\mathrm{th}}$,
where $H^{\prime}(Fr)$ is a scaling function.
This equation is qualitatively consistent with the scaling law \eqref{eq:h_Fr} for $h$. 
This indicates that the surface flow layer thickness shows the same scaling regardless of the definition of $h$.

\section{Surface flow layer thickness $h$ for $Fr \ll 1$}
\label{appG}
In this section, we show $H^{(0)}>0$ in Eq. \eqref{eq:H_expand}.
The velocity $u(z)$ is a decreasing function of $z$ and $u(z)=0$ at $z=h$.
The volumetric flow rates in the surface flow layer and the static flow regime are given by
\begin{equation}
    Q_+ = \int_0^h dz \ u(z) ,
    \label{eq:Q_+}
\end{equation}
and 
\begin{equation}
    Q_- = -\int_h^{D/2} dz \ u(z) ,
    \label{eq:Q_-}
\end{equation}
respectively.
Since the system is in a steady state, these volumetric flow rates satisfy
\begin{equation}
    Q_+ = Q_- .
    \label{eq:Q_conservation}
\end{equation}

The velocity $u(z)$ satisfies Eq. \eqref{eq:u_Fr} and its scaling function $U( \tilde z: Fr)$ is a decreasing function of $\tilde z = z/D$ with $U( \tilde z=H(Fr); Fr)=0$.
Therefore, $U( \tilde z: Fr)$ is represented as
\begin{eqnarray}
U( \tilde z: Fr) =
\left\{
\begin{array}{ll}
U_+(H(Fr)-\tilde z; Fr)& \tilde z < H(Fr) \\
 -U_-(\tilde z - H(Fr); Fr)&\tilde z \ge H(Fr) 
\end{array}
\label{eq:Upm}
\right. ,
\end{eqnarray}
where $U_\pm(\zeta ; Fr)$ is an increasing function of $\zeta$ satisfying
\begin{equation}
U_\pm(\zeta=0 ; Fr)=0, \quad U_\pm(\zeta>0 ; Fr)>0 .
\label{eq:Upm_zeta_dependence}
\end{equation}
Substituting Eqs. \eqref{eq:u_Fr} and \eqref{eq:Upm} into Eqs. \eqref{eq:Q_+} and \eqref{eq:Q_-}, $Q_\pm$ is expressed as
\begin{eqnarray}
    Q_+ & = &\Omega D^2 \int_0^{H(Fr)} d\zeta \ U_+(\zeta; Fr),
    \label{eq:Q_+_rewrite} \\
    Q_- & = &\Omega D^2 \int_0^{\frac{1}{2} - H(Fr)} d\zeta \ U_-(\zeta; Fr).
    \label{eq:Q_-_rewrite}
\end{eqnarray}

Since $Fr\ll 1$, $U_\pm(\zeta ; Fr)$ is expanded as follows:
\begin{equation}
\begin{array}{ll}
U_\pm(\zeta; Fr) = U_\pm^{(0)}(\zeta) + Fr U_\pm^{(1)}(\zeta) + O(Fr^2) ,
\end{array}
\label{eq:Upm_expand}
\end{equation}
where $U_\pm^{(0)}(\zeta)$ and $U_\pm^{(1)}(\zeta)$ are functions depending only on $\zeta$.
Considering $U_\pm(\zeta; Fr)$ in the limit of $Fr\to 0$ and Eq. \eqref{eq:Upm_zeta_dependence}, we find that $U_\pm^{(0)}(\zeta)$ satisfies
\begin{equation}
U_\pm^{(0)}(\zeta=0 )=0, \quad U_\pm^{(0)}(\zeta>0)>0 .
\label{eq:U0pm_zeta_dependence}
\end{equation}
Substituting Eqs. \eqref{eq:H_expand}, \eqref{eq:Q_+_rewrite}, \eqref{eq:Q_-_rewrite}, and \eqref{eq:Upm_expand} into Eq. \eqref{eq:Q_conservation}, we obtain
\begin{align}
     \left \{ \int_0^{H^{(0)}} d\zeta \ U_+^{(0)}(\zeta) -  \int_0^{\frac{1}{2} - H^{(0)}} d\zeta \ U_-^{(0)}(\zeta) \right \} + O(Fr) = 0 .
    \label{eq:Int_Upm_expand}
\end{align}
Since this equation holds for any $Fr$, we obtain
\begin{eqnarray}
     \int_0^{H^{(0)}} d\zeta \ U_+^{(0)}(\zeta) =  \int_0^{\frac{1}{2} - H^{(0)}} d\zeta \ U_-^{(0)}(\zeta).
    \label{eq:Upm_equation}
\end{eqnarray}
Since $ 0 \le h/D \le 1/2$ from the definition, $H^{(0)}$ satisfies $0 \le H^{(0)} \le 1/2$.
Here, $H^{(0)} $ is the solution of Eq. \eqref{eq:Upm_equation}; however, $H^{(0)}=0$ does not satisfy this equation due to Eq. \eqref{eq:Upm_zeta_dependence}. 
Therefore, we obtain
\begin{equation}
H^{(0)}>0 .
\label{eq:H0}
\end{equation}
This equation with Eqs. \eqref{eq:H_expand} and \eqref{eq:h_Fr} explains the behavior of $h$ in Fig. \ref{fig:Fr-h}.

\bibliography{ref}% Produces the bibliography via BibTeX.

%aipnum4-2.bst 2019-01-14 (MD) hand-edited version of apsrev4-1.bst
\begin{thebibliography}{52}%
\makeatletter
\providecommand \@ifxundefined [1]{%
 \@ifx{#1\undefined}
}%
\providecommand \@ifnum [1]{%
 \ifnum #1\expandafter \@firstoftwo
 \else \expandafter \@secondoftwo
 \fi
}%
\providecommand \@ifx [1]{%
 \ifx #1\expandafter \@firstoftwo
 \else \expandafter \@secondoftwo
 \fi
}%
\providecommand \natexlab [1]{#1}%
\providecommand \enquote  [1]{``#1''}%
\providecommand \bibnamefont  [1]{#1}%
\providecommand \bibfnamefont [1]{#1}%
\providecommand \citenamefont [1]{#1}%
\providecommand \href@noop [0]{\@secondoftwo}%
\providecommand \href [0]{\begingroup \@sanitize@url \@href}%
\providecommand \@href[1]{\@@startlink{#1}\@@href}%
\providecommand \@@href[1]{\endgroup#1\@@endlink}%
\providecommand \@sanitize@url [0]{\catcode `\\12\catcode `\$12\catcode
  `\&12\catcode `\#12\catcode `\^12\catcode `\_12\catcode `\%12\relax}%
\providecommand \@@startlink[1]{}%
\providecommand \@@endlink[0]{}%
\providecommand \url  [0]{\begingroup\@sanitize@url \@url }%
\providecommand \@url [1]{\endgroup\@href {#1}{\urlprefix }}%
\providecommand \urlprefix  [0]{URL }%
\providecommand \Eprint [0]{\href }%
\providecommand \doibase [0]{https://doi.org/}%
\providecommand \selectlanguage [0]{\@gobble}%
\providecommand \bibinfo  [0]{\@secondoftwo}%
\providecommand \bibfield  [0]{\@secondoftwo}%
\providecommand \translation [1]{[#1]}%
\providecommand \BibitemOpen [0]{}%
\providecommand \bibitemStop [0]{}%
\providecommand \bibitemNoStop [0]{.\EOS\space}%
\providecommand \EOS [0]{\spacefactor3000\relax}%
\providecommand \BibitemShut  [1]{\csname bibitem#1\endcsname}%
\let\auto@bib@innerbib\@empty
%</preamble>
\bibitem [{\citenamefont {Mellmann}(2001)}]{Mellmann2001}%
  \BibitemOpen
  \bibfield  {author} {\bibinfo {author} {\bibfnamefont {J.}~\bibnamefont
  {Mellmann}},\ }\bibfield  {title} {\enquote {\bibinfo {title} {The transverse
  motion of solids in rotating cylinders-forms of motion and transition
  behavior},}\ }\href
  {https://doi.org/https://doi.org/10.1016/k0032-5910(00)00402-2} {\bibfield
  {journal} {\bibinfo  {journal} {Powder technology}\ }\textbf {\bibinfo
  {volume} {118}},\ \bibinfo {pages} {251--270} (\bibinfo {year}
  {2001})}\BibitemShut {NoStop}%
\bibitem [{\citenamefont {Parker}\ \emph {et~al.}(1997)\citenamefont {Parker},
  \citenamefont {Dijkstra}, \citenamefont {Martin},\ and\ \citenamefont
  {Seville}}]{Parker1997}%
  \BibitemOpen
  \bibfield  {author} {\bibinfo {author} {\bibfnamefont {D.}~\bibnamefont
  {Parker}}, \bibinfo {author} {\bibfnamefont {A.}~\bibnamefont {Dijkstra}},
  \bibinfo {author} {\bibfnamefont {T.}~\bibnamefont {Martin}},\ and\ \bibinfo
  {author} {\bibfnamefont {J.}~\bibnamefont {Seville}},\ }\bibfield  {title}
  {\enquote {\bibinfo {title} {Positron emission particle tracking studies of
  spherical particle motion in rotating drums},}\ }\href
  {https://doi.org/https://doi.org/10.1016/S0009-2509(97)00030-4} {\bibfield
  {journal} {\bibinfo  {journal} {Chem. Eng. Sci.}\ }\textbf {\bibinfo {volume}
  {52}},\ \bibinfo {pages} {2011--2022} (\bibinfo {year} {1997})}\BibitemShut
  {NoStop}%
\bibitem [{\citenamefont {Khakhar}\ \emph {et~al.}(1997)\citenamefont
  {Khakhar}, \citenamefont {McCarthy}, \citenamefont {Shinbrot},\ and\
  \citenamefont {Ottino}}]{Khakhar1997}%
  \BibitemOpen
  \bibfield  {author} {\bibinfo {author} {\bibfnamefont {D.~V.}\ \bibnamefont
  {Khakhar}}, \bibinfo {author} {\bibfnamefont {J.~J.}\ \bibnamefont
  {McCarthy}}, \bibinfo {author} {\bibfnamefont {T.}~\bibnamefont {Shinbrot}},\
  and\ \bibinfo {author} {\bibfnamefont {J.~M.}\ \bibnamefont {Ottino}},\
  }\bibfield  {title} {\enquote {\bibinfo {title} {Transverse flow and mixing
  of granular materials in a rotating cylinder},}\ }\href
  {https://doi.org/10.1063/1.869172} {\bibfield  {journal} {\bibinfo  {journal}
  {Phys. Fluids}\ }\textbf {\bibinfo {volume} {9}},\ \bibinfo {pages} {31}
  (\bibinfo {year} {1997})}\BibitemShut {NoStop}%
\bibitem [{\citenamefont {Jain}, \citenamefont {Ottino},\ and\ \citenamefont
  {Lueptow}(2004)}]{Jain2004}%
  \BibitemOpen
  \bibfield  {author} {\bibinfo {author} {\bibfnamefont {N.}~\bibnamefont
  {Jain}}, \bibinfo {author} {\bibfnamefont {J.~M.}\ \bibnamefont {Ottino}},\
  and\ \bibinfo {author} {\bibfnamefont {R.~M.}\ \bibnamefont {Lueptow}},\
  }\bibfield  {title} {\enquote {\bibinfo {title} {Effect of interstitial fluid
  on a granular flowing layer},}\ }\href
  {https://doi.org/10.1017/S0022112004008869} {\bibfield  {journal} {\bibinfo
  {journal} {J. Fluid Mech.}\ }\textbf {\bibinfo {volume} {508}},\ \bibinfo
  {pages} {23--44} (\bibinfo {year} {2004})}\BibitemShut {NoStop}%
\bibitem [{\citenamefont {F{\'e}lix}, \citenamefont {Falk},\ and\ \citenamefont
  {d'Ortona}(2007)}]{Felix2007}%
  \BibitemOpen
  \bibfield  {author} {\bibinfo {author} {\bibfnamefont {G.}~\bibnamefont
  {F{\'e}lix}}, \bibinfo {author} {\bibfnamefont {V.}~\bibnamefont {Falk}},\
  and\ \bibinfo {author} {\bibfnamefont {U.}~\bibnamefont {d'Ortona}},\
  }\bibfield  {title} {\enquote {\bibinfo {title} {Granular flows in a rotating
  drum: the scaling law between velocity and thickness of the flow},}\ }\href
  {https://doi.org/https://doi.org/10.1140/epje/e2007-00002-5} {\bibfield
  {journal} {\bibinfo  {journal} {Eur. Phys. J. E.}\ }\textbf {\bibinfo
  {volume} {22}},\ \bibinfo {pages} {25--31} (\bibinfo {year}
  {2007})}\BibitemShut {NoStop}%
\bibitem [{\citenamefont {Orpe}\ and\ \citenamefont
  {Khakhar}(2007)}]{Orpe2007}%
  \BibitemOpen
  \bibfield  {author} {\bibinfo {author} {\bibfnamefont {A.~V.}\ \bibnamefont
  {Orpe}}\ and\ \bibinfo {author} {\bibfnamefont {D.~V.}\ \bibnamefont
  {Khakhar}},\ }\bibfield  {title} {\enquote {\bibinfo {title} {Rheology of
  surface granular flows},}\ }\href {https://doi.org/10.1017/S002211200600320X}
  {\bibfield  {journal} {\bibinfo  {journal} {J. Fluid Mech.}\ }\textbf
  {\bibinfo {volume} {571}},\ \bibinfo {pages} {1} (\bibinfo {year}
  {2007})}\BibitemShut {NoStop}%
\bibitem [{\citenamefont {Chen}\ \emph {et~al.}(2024)\citenamefont {Chen},
  \citenamefont {Suo}, \citenamefont {Dong}, \citenamefont {Zhong},
  \citenamefont {Wei},\ and\ \citenamefont {Gan}}]{Chen2024}%
  \BibitemOpen
  \bibfield  {author} {\bibinfo {author} {\bibfnamefont {Y.}~\bibnamefont
  {Chen}}, \bibinfo {author} {\bibfnamefont {S.}~\bibnamefont {Suo}}, \bibinfo
  {author} {\bibfnamefont {M.}~\bibnamefont {Dong}}, \bibinfo {author}
  {\bibfnamefont {H.}~\bibnamefont {Zhong}}, \bibinfo {author} {\bibfnamefont
  {D.}~\bibnamefont {Wei}},\ and\ \bibinfo {author} {\bibfnamefont
  {Y.}~\bibnamefont {Gan}},\ }\bibfield  {title} {\enquote {\bibinfo {title}
  {Effects of particle density and fluid properties on mono-dispersed granular
  flows in a rotating drum},}\ }\href {https://doi.org/10.1063/5.0229006}
  {\bibfield  {journal} {\bibinfo  {journal} {Phys. Fluids}\ }\textbf {\bibinfo
  {volume} {36}},\ \bibinfo {pages} {103324} (\bibinfo {year}
  {2024})}\BibitemShut {NoStop}%
\bibitem [{\citenamefont {Pignatel}\ \emph {et~al.}(2012)\citenamefont
  {Pignatel}, \citenamefont {Asselin}, \citenamefont {Krieger}, \citenamefont
  {Christov}, \citenamefont {Ottino},\ and\ \citenamefont
  {Lueptow}}]{Pignatel2012}%
  \BibitemOpen
  \bibfield  {author} {\bibinfo {author} {\bibfnamefont {F.}~\bibnamefont
  {Pignatel}}, \bibinfo {author} {\bibfnamefont {C.}~\bibnamefont {Asselin}},
  \bibinfo {author} {\bibfnamefont {L.}~\bibnamefont {Krieger}}, \bibinfo
  {author} {\bibfnamefont {I.~C.}\ \bibnamefont {Christov}}, \bibinfo {author}
  {\bibfnamefont {J.~M.}\ \bibnamefont {Ottino}},\ and\ \bibinfo {author}
  {\bibfnamefont {R.~M.}\ \bibnamefont {Lueptow}},\ }\bibfield  {title}
  {\enquote {\bibinfo {title} {Parameters and scalings for dry and immersed
  granular flowing layers in rotating tumblers},}\ }\href
  {https://doi.org/10.1103/PhysRevE.86.011304} {\bibfield  {journal} {\bibinfo
  {journal} {Phys. Rev. E}\ }\textbf {\bibinfo {volume} {86}},\ \bibinfo
  {pages} {011304} (\bibinfo {year} {2012})}\BibitemShut {NoStop}%
\bibitem [{\citenamefont {Yamane}\ \emph {et~al.}(1998)\citenamefont {Yamane},
  \citenamefont {Nakagawa}, \citenamefont {Altobelli}, \citenamefont {Tanaka},\
  and\ \citenamefont {Tsuji}}]{Yamane1997}%
  \BibitemOpen
  \bibfield  {author} {\bibinfo {author} {\bibfnamefont {K.}~\bibnamefont
  {Yamane}}, \bibinfo {author} {\bibfnamefont {M.}~\bibnamefont {Nakagawa}},
  \bibinfo {author} {\bibfnamefont {S.~A.}\ \bibnamefont {Altobelli}}, \bibinfo
  {author} {\bibfnamefont {T.}~\bibnamefont {Tanaka}},\ and\ \bibinfo {author}
  {\bibfnamefont {Y.}~\bibnamefont {Tsuji}},\ }\bibfield  {title} {\enquote
  {\bibinfo {title} {Steady particulate flows in a horizontal rotating
  cylinder},}\ }\href {https://doi.org/https://doi.org/10.1063/1.869858}
  {\bibfield  {journal} {\bibinfo  {journal} {Phys. Fluids}\ }\textbf {\bibinfo
  {volume} {10}},\ \bibinfo {pages} {1419--1427} (\bibinfo {year}
  {1998})}\BibitemShut {NoStop}%
\bibitem [{\citenamefont {Dury}\ \emph {et~al.}(1998)\citenamefont {Dury},
  \citenamefont {Ristow}, \citenamefont {Moss},\ and\ \citenamefont
  {Nakagawa}}]{Dury1998}%
  \BibitemOpen
  \bibfield  {author} {\bibinfo {author} {\bibfnamefont {C.}~\bibnamefont
  {Dury}}, \bibinfo {author} {\bibfnamefont {G.}~\bibnamefont {Ristow}},
  \bibinfo {author} {\bibfnamefont {J.}~\bibnamefont {Moss}},\ and\ \bibinfo
  {author} {\bibfnamefont {M.}~\bibnamefont {Nakagawa}},\ }\bibfield  {title}
  {\enquote {\bibinfo {title} {Boundary effects on the angle of repose in
  rotating cylinders},}\ }\href {https://doi.org/10.1103/PhysRevE.57.4491}
  {\bibfield  {journal} {\bibinfo  {journal} {Phys. Rev. E}\ }\textbf {\bibinfo
  {volume} {57}},\ \bibinfo {pages} {4491--4497} (\bibinfo {year}
  {1998})}\BibitemShut {NoStop}%
\bibitem [{\citenamefont {Renouf}\ \emph {et~al.}(2005)\citenamefont {Renouf},
  \citenamefont {Bonamy}, \citenamefont {Dubois},\ and\ \citenamefont
  {Alart}}]{Renouf2005}%
  \BibitemOpen
  \bibfield  {author} {\bibinfo {author} {\bibfnamefont {M.}~\bibnamefont
  {Renouf}}, \bibinfo {author} {\bibfnamefont {D.}~\bibnamefont {Bonamy}},
  \bibinfo {author} {\bibfnamefont {F.}~\bibnamefont {Dubois}},\ and\ \bibinfo
  {author} {\bibfnamefont {P.}~\bibnamefont {Alart}},\ }\bibfield  {title}
  {\enquote {\bibinfo {title} {{Numerical simulation of two-dimensional steady
  granular flows in rotating drum: On surface flow rheology}},}\ }\href
  {https://doi.org/10.1063/1.2063347} {\bibfield  {journal} {\bibinfo
  {journal} {Phys. Fluids}\ }\textbf {\bibinfo {volume} {17}},\ \bibinfo
  {pages} {103303} (\bibinfo {year} {2005})}\BibitemShut {NoStop}%
\bibitem [{\citenamefont {Yang}, \citenamefont {Zou},\ and\ \citenamefont
  {Yu}(2003)}]{Yang2003}%
  \BibitemOpen
  \bibfield  {author} {\bibinfo {author} {\bibfnamefont {R.}~\bibnamefont
  {Yang}}, \bibinfo {author} {\bibfnamefont {R.}~\bibnamefont {Zou}},\ and\
  \bibinfo {author} {\bibfnamefont {A.}~\bibnamefont {Yu}},\ }\bibfield
  {title} {\enquote {\bibinfo {title} {Microdynamic analysis of particle flow
  in a horizontal rotating drum},}\ }\href
  {https://doi.org/https://doi.org/10.1016/S0032-5910(02)00257-7} {\bibfield
  {journal} {\bibinfo  {journal} {Powder Technol.}\ }\textbf {\bibinfo {volume}
  {130}},\ \bibinfo {pages} {138--146} (\bibinfo {year} {2003})}\BibitemShut
  {NoStop}%
\bibitem [{\citenamefont {Yang}\ \emph {et~al.}(2008)\citenamefont {Yang},
  \citenamefont {Yu}, \citenamefont {McElroy},\ and\ \citenamefont
  {Bao}}]{Yang2008}%
  \BibitemOpen
  \bibfield  {author} {\bibinfo {author} {\bibfnamefont {R.}~\bibnamefont
  {Yang}}, \bibinfo {author} {\bibfnamefont {A.}~\bibnamefont {Yu}}, \bibinfo
  {author} {\bibfnamefont {L.}~\bibnamefont {McElroy}},\ and\ \bibinfo {author}
  {\bibfnamefont {J.}~\bibnamefont {Bao}},\ }\bibfield  {title} {\enquote
  {\bibinfo {title} {Numerical simulation of particle dynamics in different
  flow regimes in a rotating drum},}\ }\href
  {https://doi.org/https://doi.org/10.1016/j.powtec.2008.04.081} {\bibfield
  {journal} {\bibinfo  {journal} {Powder Technol.}\ }\textbf {\bibinfo {volume}
  {188}},\ \bibinfo {pages} {170--177} (\bibinfo {year} {2008})}\BibitemShut
  {NoStop}%
\bibitem [{\citenamefont {Da~Cruz}\ \emph {et~al.}(2005)\citenamefont
  {Da~Cruz}, \citenamefont {Emam}, \citenamefont {Prochnow}, \citenamefont
  {Roux},\ and\ \citenamefont {Chevoir}}]{DaCruiz2005}%
  \BibitemOpen
  \bibfield  {author} {\bibinfo {author} {\bibfnamefont {F.}~\bibnamefont
  {Da~Cruz}}, \bibinfo {author} {\bibfnamefont {S.}~\bibnamefont {Emam}},
  \bibinfo {author} {\bibfnamefont {M.}~\bibnamefont {Prochnow}}, \bibinfo
  {author} {\bibfnamefont {J.}~\bibnamefont {Roux}},\ and\ \bibinfo {author}
  {\bibfnamefont {F.}~\bibnamefont {Chevoir}},\ }\bibfield  {title} {\enquote
  {\bibinfo {title} {Rheophysics of dense granular materials: Discrete
  simulation of plane shear flows},}\ }\href
  {https://doi.org/https://doi.org/10.1103/PhysRevE.72.021309} {\bibfield
  {journal} {\bibinfo  {journal} {Phys. Rev. E}\ }\textbf {\bibinfo {volume}
  {72}},\ \bibinfo {pages} {021309} (\bibinfo {year} {2005})}\BibitemShut
  {NoStop}%
\bibitem [{\citenamefont {Jop}, \citenamefont {Forterre},\ and\ \citenamefont
  {Pouliquen}(2006)}]{Jop2006}%
  \BibitemOpen
  \bibfield  {author} {\bibinfo {author} {\bibfnamefont {P.}~\bibnamefont
  {Jop}}, \bibinfo {author} {\bibfnamefont {Y.}~\bibnamefont {Forterre}},\ and\
  \bibinfo {author} {\bibfnamefont {O.}~\bibnamefont {Pouliquen}},\ }\bibfield
  {title} {\enquote {\bibinfo {title} {{A constitutive law for dense granular
  flows}},}\ }\href {https://doi.org/10.1038/nature04801} {\bibfield  {journal}
  {\bibinfo  {journal} {Nature}\ }\textbf {\bibinfo {volume} {441}},\ \bibinfo
  {pages} {727--730} (\bibinfo {year} {2006})}\BibitemShut {NoStop}%
\bibitem [{\citenamefont {Peyneau}\ and\ \citenamefont
  {Roux}(2008)}]{Peyneau2008}%
  \BibitemOpen
  \bibfield  {author} {\bibinfo {author} {\bibfnamefont {P.-E.}\ \bibnamefont
  {Peyneau}}\ and\ \bibinfo {author} {\bibfnamefont {J.-N.}\ \bibnamefont
  {Roux}},\ }\bibfield  {title} {\enquote {\bibinfo {title} {Frictionless bead
  packs have macroscopic friction, but no dilatancy},}\ }\href
  {https://doi.org/10.1103/PhysRevE.78.011307} {\bibfield  {journal} {\bibinfo
  {journal} {Phys. Rev. E}\ }\textbf {\bibinfo {volume} {78}},\ \bibinfo
  {pages} {011307} (\bibinfo {year} {2008})}\BibitemShut {NoStop}%
\bibitem [{\citenamefont {Kamrin}\ and\ \citenamefont
  {Koval}(2012)}]{Kamrin2012}%
  \BibitemOpen
  \bibfield  {author} {\bibinfo {author} {\bibfnamefont {K.}~\bibnamefont
  {Kamrin}}\ and\ \bibinfo {author} {\bibfnamefont {G.}~\bibnamefont {Koval}},\
  }\bibfield  {title} {\enquote {\bibinfo {title} {Nonlocal constitutive
  relation for steady granular flow},}\ }\href
  {https://doi.org/https://doi.org/10.1103/PhysRevLett.108.178301} {\bibfield
  {journal} {\bibinfo  {journal} {Phys. Rev. Lett.}\ }\textbf {\bibinfo
  {volume} {108}},\ \bibinfo {pages} {178301} (\bibinfo {year}
  {2012})}\BibitemShut {NoStop}%
\bibitem [{\citenamefont {Heyman}\ \emph {et~al.}(2017)\citenamefont {Heyman},
  \citenamefont {Delannay}, \citenamefont {Tabuteau},\ and\ \citenamefont
  {Valance}}]{Heyman2017}%
  \BibitemOpen
  \bibfield  {author} {\bibinfo {author} {\bibfnamefont {J.}~\bibnamefont
  {Heyman}}, \bibinfo {author} {\bibfnamefont {R.}~\bibnamefont {Delannay}},
  \bibinfo {author} {\bibfnamefont {H.}~\bibnamefont {Tabuteau}},\ and\
  \bibinfo {author} {\bibfnamefont {A.}~\bibnamefont {Valance}},\ }\bibfield
  {title} {\enquote {\bibinfo {title} {Compressibility regularizes the
  $\mu(i)$-rheology for dense granular flows},}\ }\href
  {https://doi.org/https://doi.org/10.1017/jfm.2017.612} {\bibfield  {journal}
  {\bibinfo  {journal} {J. Fluid Mech.}\ }\textbf {\bibinfo {volume} {830}},\
  \bibinfo {pages} {553--568} (\bibinfo {year} {2017})}\BibitemShut {NoStop}%
\bibitem [{\citenamefont {Kim}\ and\ \citenamefont {Kamrin}(2020)}]{Kim2020}%
  \BibitemOpen
  \bibfield  {author} {\bibinfo {author} {\bibfnamefont {S.}~\bibnamefont
  {Kim}}\ and\ \bibinfo {author} {\bibfnamefont {K.}~\bibnamefont {Kamrin}},\
  }\bibfield  {title} {\enquote {\bibinfo {title} {Power-law scaling in
  granular rheology across flow geometries},}\ }\href
  {https://doi.org/https://doi.org/10.1103/PhysRevLett.125.088002} {\bibfield
  {journal} {\bibinfo  {journal} {Phys. Rev. Lett.}\ }\textbf {\bibinfo
  {volume} {125}},\ \bibinfo {pages} {088002} (\bibinfo {year}
  {2020})}\BibitemShut {NoStop}%
\bibitem [{\citenamefont {MiDi}(2004)}]{G.D.R.MiDi2004}%
  \BibitemOpen
  \bibfield  {author} {\bibinfo {author} {\bibfnamefont {G.}~\bibnamefont
  {MiDi}},\ }\bibfield  {title} {\enquote {\bibinfo {title} {On dense granular
  flows},}\ }\href {https://doi.org/https://doi.org/10.1140/epje/i2003-10153-0}
  {\bibfield  {journal} {\bibinfo  {journal} {Eur. Phys. J. E.}\ }\textbf
  {\bibinfo {volume} {14}},\ \bibinfo {pages} {341--365} (\bibinfo {year}
  {2004})}\BibitemShut {NoStop}%
\bibitem [{\citenamefont {Zheng}\ and\ \citenamefont {Yu}(2015)}]{Zheng2015}%
  \BibitemOpen
  \bibfield  {author} {\bibinfo {author} {\bibfnamefont {Q.}~\bibnamefont
  {Zheng}}\ and\ \bibinfo {author} {\bibfnamefont {A.}~\bibnamefont {Yu}},\
  }\bibfield  {title} {\enquote {\bibinfo {title} {Modelling the granular flow
  in a rotating drum by the eulerian finite element method},}\ }\href
  {https://doi.org/https://doi.org/10.1016/j.powtec.2015.08.025} {\bibfield
  {journal} {\bibinfo  {journal} {Powder Technol.}\ }\textbf {\bibinfo {volume}
  {286}},\ \bibinfo {pages} {361--370} (\bibinfo {year} {2015})}\BibitemShut
  {NoStop}%
\bibitem [{\citenamefont {Zheng}\ \emph {et~al.}(2019)\citenamefont {Zheng},
  \citenamefont {Bai}, \citenamefont {Yang},\ and\ \citenamefont
  {Yu}}]{Zheng2019}%
  \BibitemOpen
  \bibfield  {author} {\bibinfo {author} {\bibfnamefont {Q.}~\bibnamefont
  {Zheng}}, \bibinfo {author} {\bibfnamefont {L.}~\bibnamefont {Bai}}, \bibinfo
  {author} {\bibfnamefont {L.}~\bibnamefont {Yang}},\ and\ \bibinfo {author}
  {\bibfnamefont {A.}~\bibnamefont {Yu}},\ }\bibfield  {title} {\enquote
  {\bibinfo {title} {110th anniversary: Continuum modeling of granular mixing
  in a rotating drum},}\ }\href
  {https://doi.org/https://doi.org/10.1021/acs.iecr.9b03642} {\bibfield
  {journal} {\bibinfo  {journal} {Ind. Eng. Chem. Res}\ }\textbf {\bibinfo
  {volume} {58}},\ \bibinfo {pages} {19251--19262} (\bibinfo {year}
  {2019})}\BibitemShut {NoStop}%
\bibitem [{\citenamefont {Liu}, \citenamefont {Gonzalez},\ and\ \citenamefont
  {Wassgren}(2018)}]{Liu2018}%
  \BibitemOpen
  \bibfield  {author} {\bibinfo {author} {\bibfnamefont {Y.}~\bibnamefont
  {Liu}}, \bibinfo {author} {\bibfnamefont {M.}~\bibnamefont {Gonzalez}},\ and\
  \bibinfo {author} {\bibfnamefont {C.}~\bibnamefont {Wassgren}},\ }\bibfield
  {title} {\enquote {\bibinfo {title} {Modeling granular material blending in a
  rotating drum using a finite element method and advection-diffusion equation
  multiscale model},}\ }\href
  {https://doi.org/https://doi.org/10.1002/aic.16179} {\bibfield  {journal}
  {\bibinfo  {journal} {AIChE J.}\ }\textbf {\bibinfo {volume} {64}},\ \bibinfo
  {pages} {3277--3292} (\bibinfo {year} {2018})}\BibitemShut {NoStop}%
\bibitem [{\citenamefont {Arseni}\ \emph {et~al.}(2020)\citenamefont {Arseni},
  \citenamefont {De~Monaco}, \citenamefont {Greco},\ and\ \citenamefont
  {Maffettone}}]{Arseni2020}%
  \BibitemOpen
  \bibfield  {author} {\bibinfo {author} {\bibfnamefont {A.}~\bibnamefont
  {Arseni}}, \bibinfo {author} {\bibfnamefont {G.}~\bibnamefont {De~Monaco}},
  \bibinfo {author} {\bibfnamefont {F.}~\bibnamefont {Greco}},\ and\ \bibinfo
  {author} {\bibfnamefont {P.}~\bibnamefont {Maffettone}},\ }\bibfield  {title}
  {\enquote {\bibinfo {title} {Granular flow in rotating drums through
  simulations adopting a continuum constitutive equation},}\ }\href
  {https://doi.org/10.1063/5.0018694} {\bibfield  {journal} {\bibinfo
  {journal} {Phys. Fluids}\ }\textbf {\bibinfo {volume} {32}},\ \bibinfo
  {pages} {093305} (\bibinfo {year} {2020})}\BibitemShut {NoStop}%
\bibitem [{\citenamefont {Santos}\ \emph {et~al.}(2013)\citenamefont {Santos},
  \citenamefont {Petri}, \citenamefont {Duarte},\ and\ \citenamefont
  {Barrozo}}]{Santos2013}%
  \BibitemOpen
  \bibfield  {author} {\bibinfo {author} {\bibfnamefont {D.}~\bibnamefont
  {Santos}}, \bibinfo {author} {\bibfnamefont {I.}~\bibnamefont {Petri}},
  \bibinfo {author} {\bibfnamefont {C.}~\bibnamefont {Duarte}},\ and\ \bibinfo
  {author} {\bibfnamefont {M.}~\bibnamefont {Barrozo}},\ }\bibfield  {title}
  {\enquote {\bibinfo {title} {Experimental and cfd study of the hydrodynamic
  behavior in a rotating drum},}\ }\href
  {https://doi.org/https://doi.org/10.1016/j.powtec.2013.10.003} {\bibfield
  {journal} {\bibinfo  {journal} {Powder Technol.}\ }\textbf {\bibinfo {volume}
  {250}},\ \bibinfo {pages} {52--62} (\bibinfo {year} {2013})}\BibitemShut
  {NoStop}%
\bibitem [{\citenamefont {Batchelor}(2000)}]{Batchelor2000}%
  \BibitemOpen
  \bibfield  {author} {\bibinfo {author} {\bibfnamefont {G.~K.}\ \bibnamefont
  {Batchelor}},\ }\href {https://doi.org/10.1017/CBO9780511800955} {\emph
  {\bibinfo {title} {An Introduction to Fluid Dynamics}}}\ (\bibinfo
  {publisher} {Cambridge University Press},\ \bibinfo {year}
  {2000})\BibitemShut {NoStop}%
\bibitem [{\citenamefont {Cundall}\ and\ \citenamefont
  {Strack}(1979)}]{Cundall1979}%
  \BibitemOpen
  \bibfield  {author} {\bibinfo {author} {\bibfnamefont {P.}~\bibnamefont
  {Cundall}}\ and\ \bibinfo {author} {\bibfnamefont {O.}~\bibnamefont
  {Strack}},\ }\bibfield  {title} {\enquote {\bibinfo {title} {A discrete
  numerical model for granular assemblies},}\ }\href
  {https://doi.org/10.1680/geot.1979.29.1.47} {\bibfield  {journal} {\bibinfo
  {journal} {G{\'e}otechnique}\ }\textbf {\bibinfo {volume} {29}},\ \bibinfo
  {pages} {47--65} (\bibinfo {year} {1979})}\BibitemShut {NoStop}%
\bibitem [{\citenamefont {Sh\"{a}fer}, \citenamefont {Dippel},\ and\
  \citenamefont {Wolf}(1996)}]{Shafer1996}%
  \BibitemOpen
  \bibfield  {author} {\bibinfo {author} {\bibfnamefont {J.}~\bibnamefont
  {Sh\"{a}fer}}, \bibinfo {author} {\bibfnamefont {S.}~\bibnamefont {Dippel}},\
  and\ \bibinfo {author} {\bibfnamefont {D.}~\bibnamefont {Wolf}},\ }\bibfield
  {title} {\enquote {\bibinfo {title} {Force schemes in simulations of granular
  materials},}\ }\href {https://doi.org/https://doi.org/10.1051/jp1:1996129}
  {\bibfield  {journal} {\bibinfo  {journal} {J. Phys. I France}\ }\textbf
  {\bibinfo {volume} {6}},\ \bibinfo {pages} {5--20} (\bibinfo {year}
  {1996})}\BibitemShut {NoStop}%
\bibitem [{\citenamefont {Campbell}(2002)}]{Campbell2002}%
  \BibitemOpen
  \bibfield  {author} {\bibinfo {author} {\bibfnamefont {C.}~\bibnamefont
  {Campbell}},\ }\bibfield  {title} {\enquote {\bibinfo {title} {Granular shear
  flows at the elastic limit},}\ }\href
  {https://doi.org/10.1017/S002211200200109X} {\bibfield  {journal} {\bibinfo
  {journal} {J. Fluid Mech.}\ }\textbf {\bibinfo {volume} {465}},\ \bibinfo
  {pages} {261--291} (\bibinfo {year} {2002})}\BibitemShut {NoStop}%
\bibitem [{\citenamefont {Silbert}\ \emph {et~al.}(2001)\citenamefont
  {Silbert}, \citenamefont {D.}, \citenamefont {Grest}, \citenamefont {Halsey},
  \citenamefont {Levine},\ and\ \citenamefont {Plimpton}}]{Silbert2001}%
  \BibitemOpen
  \bibfield  {author} {\bibinfo {author} {\bibfnamefont {L.}~\bibnamefont
  {Silbert}}, \bibinfo {author} {\bibfnamefont {E.}~\bibnamefont {D.}},
  \bibinfo {author} {\bibfnamefont {G.}~\bibnamefont {Grest}}, \bibinfo
  {author} {\bibfnamefont {T.}~\bibnamefont {Halsey}}, \bibinfo {author}
  {\bibfnamefont {D.}~\bibnamefont {Levine}},\ and\ \bibinfo {author}
  {\bibfnamefont {S.}~\bibnamefont {Plimpton}},\ }\bibfield  {title} {\enquote
  {\bibinfo {title} {Granular flow down an inclined plane: Bagnold scaling and
  rheology},}\ }\href
  {https://doi.org/https://doi.org/10.1103/PhysRevE.64.051302} {\bibfield
  {journal} {\bibinfo  {journal} {Phys. Rev. E}\ }\textbf {\bibinfo {volume}
  {64}},\ \bibinfo {pages} {14} (\bibinfo {year} {2001})}\BibitemShut {NoStop}%
\bibitem [{\citenamefont {Watanabe}\ and\ \citenamefont
  {Goto}(2022)}]{Watanabe2022}%
  \BibitemOpen
  \bibfield  {author} {\bibinfo {author} {\bibfnamefont {D.}~\bibnamefont
  {Watanabe}}\ and\ \bibinfo {author} {\bibfnamefont {S.}~\bibnamefont
  {Goto}},\ }\bibfield  {title} {\enquote {\bibinfo {title} {Simple bladeless
  mixer with liquid-gas interface},}\ }\href
  {https://doi.org/10.1017/flo.2022.22} {\bibfield  {journal} {\bibinfo
  {journal} {Flow}\ }\textbf {\bibinfo {volume} {2}},\ \bibinfo {pages} {E28}
  (\bibinfo {year} {2022})}\BibitemShut {NoStop}%
\bibitem [{\citenamefont {Ding}\ \emph {et~al.}(2002)\citenamefont {Ding},
  \citenamefont {Forster}, \citenamefont {Seville},\ and\ \citenamefont
  {Parker}}]{Ding2002}%
  \BibitemOpen
  \bibfield  {author} {\bibinfo {author} {\bibfnamefont {Y.}~\bibnamefont
  {Ding}}, \bibinfo {author} {\bibfnamefont {R.}~\bibnamefont {Forster}},
  \bibinfo {author} {\bibfnamefont {J.}~\bibnamefont {Seville}},\ and\ \bibinfo
  {author} {\bibfnamefont {D.}~\bibnamefont {Parker}},\ }\bibfield  {title}
  {\enquote {\bibinfo {title} {Segregation of granular flow in the transverse
  plane of a rolling mode rotating drum},}\ }\href
  {https://doi.org/https://doi.org/10.1016/S0301-9322(01)00081-7} {\bibfield
  {journal} {\bibinfo  {journal} {Int. J. Multiph. Flow}\ }\textbf {\bibinfo
  {volume} {28}},\ \bibinfo {pages} {635--663} (\bibinfo {year}
  {2002})}\BibitemShut {NoStop}%
\bibitem [{\citenamefont {Bonamy}, \citenamefont {Daviaud},\ and\ \citenamefont
  {Laurent}(2002)}]{Bonamy2002}%
  \BibitemOpen
  \bibfield  {author} {\bibinfo {author} {\bibfnamefont {D.}~\bibnamefont
  {Bonamy}}, \bibinfo {author} {\bibfnamefont {F.}~\bibnamefont {Daviaud}},\
  and\ \bibinfo {author} {\bibfnamefont {L.}~\bibnamefont {Laurent}},\
  }\bibfield  {title} {\enquote {\bibinfo {title} {Experimental study of
  granular surface flows via a fast camera: a continuous description},}\ }\href
  {https://doi.org/https://doi.org/10.1063/1.1459720} {\bibfield  {journal}
  {\bibinfo  {journal} {Phys. Fluids}\ }\textbf {\bibinfo {volume} {14}},\
  \bibinfo {pages} {1666--1673} (\bibinfo {year} {2002})}\BibitemShut {NoStop}%
\bibitem [{\citenamefont {Cortet}\ \emph {et~al.}(2009)\citenamefont {Cortet},
  \citenamefont {Bonamy}, \citenamefont {Daviaud}, \citenamefont {Dauchot},
  \citenamefont {Dubrulle},\ and\ \citenamefont {Renouf}}]{Cortet2009}%
  \BibitemOpen
  \bibfield  {author} {\bibinfo {author} {\bibfnamefont {P.-P.}\ \bibnamefont
  {Cortet}}, \bibinfo {author} {\bibfnamefont {D.}~\bibnamefont {Bonamy}},
  \bibinfo {author} {\bibfnamefont {F.}~\bibnamefont {Daviaud}}, \bibinfo
  {author} {\bibfnamefont {O.}~\bibnamefont {Dauchot}}, \bibinfo {author}
  {\bibfnamefont {B.}~\bibnamefont {Dubrulle}},\ and\ \bibinfo {author}
  {\bibfnamefont {M.}~\bibnamefont {Renouf}},\ }\bibfield  {title} {\enquote
  {\bibinfo {title} {Relevance of visco-plastic theory in a multi-directional
  inhomogeneous granular flow},}\ }\href
  {https://doi.org/10.1209/0295-5075/88/14001} {\bibfield  {journal} {\bibinfo
  {journal} {EPL}\ }\textbf {\bibinfo {volume} {88}},\ \bibinfo {pages} {14001}
  (\bibinfo {year} {2009})}\BibitemShut {NoStop}%
\bibitem [{\citenamefont {Orpe}\ and\ \citenamefont
  {Khakhar}(2001)}]{Orpe2001}%
  \BibitemOpen
  \bibfield  {author} {\bibinfo {author} {\bibfnamefont {A.}~\bibnamefont
  {Orpe}}\ and\ \bibinfo {author} {\bibfnamefont {D.}~\bibnamefont {Khakhar}},\
  }\bibfield  {title} {\enquote {\bibinfo {title} {Scaling relations for
  granular flow in quasi-two-dimensional rotating cylinders},}\ }\href
  {https://doi.org/10.1103/PhysRevE.64.031302} {\bibfield  {journal} {\bibinfo
  {journal} {Phys. Rev. E}\ }\textbf {\bibinfo {volume} {64}},\ \bibinfo
  {pages} {031302} (\bibinfo {year} {2001})}\BibitemShut {NoStop}%
\bibitem [{\citenamefont {Agarwal}\ \emph {et~al.}(2021)\citenamefont
  {Agarwal}, \citenamefont {Karsai}, \citenamefont {Goldman},\ and\
  \citenamefont {Kamrin}}]{Agarwal2021}%
  \BibitemOpen
  \bibfield  {author} {\bibinfo {author} {\bibfnamefont {S.}~\bibnamefont
  {Agarwal}}, \bibinfo {author} {\bibfnamefont {A.}~\bibnamefont {Karsai}},
  \bibinfo {author} {\bibfnamefont {D.}~\bibnamefont {Goldman}},\ and\ \bibinfo
  {author} {\bibfnamefont {K.}~\bibnamefont {Kamrin}},\ }\bibfield  {title}
  {\enquote {\bibinfo {title} {Efficacy of simple continuum models for diverse
  granular intrusions},}\ }\href
  {https://doi.org/https://doi.org/10.1039/d1sm00130b} {\bibfield  {journal}
  {\bibinfo  {journal} {Soft Matter}\ }\textbf {\bibinfo {volume} {17}},\
  \bibinfo {pages} {7196--7209} (\bibinfo {year} {2021})}\BibitemShut {NoStop}%
\bibitem [{\citenamefont {Berzi}\ and\ \citenamefont
  {Jenkins}(2011)}]{Berzi2011}%
  \BibitemOpen
  \bibfield  {author} {\bibinfo {author} {\bibfnamefont {D.}~\bibnamefont
  {Berzi}}\ and\ \bibinfo {author} {\bibfnamefont {J.~T.}\ \bibnamefont
  {Jenkins}},\ }\bibfield  {title} {\enquote {\bibinfo {title} {Surface flows
  of inelastic spheres},}\ }\href {https://doi.org/10.1063/1.3532838}
  {\bibfield  {journal} {\bibinfo  {journal} {Phys. Fluids}\ }\textbf {\bibinfo
  {volume} {23}},\ \bibinfo {pages} {013303} (\bibinfo {year}
  {2011})}\BibitemShut {NoStop}%
\bibitem [{\citenamefont {Barker}\ \emph {et~al.}(2015)\citenamefont {Barker},
  \citenamefont {Schaeffer}, \citenamefont {Bohorquez},\ and\ \citenamefont
  {Gray}}]{Barker2015}%
  \BibitemOpen
  \bibfield  {author} {\bibinfo {author} {\bibfnamefont {T.}~\bibnamefont
  {Barker}}, \bibinfo {author} {\bibfnamefont {D.~G.}\ \bibnamefont
  {Schaeffer}}, \bibinfo {author} {\bibfnamefont {P.}~\bibnamefont
  {Bohorquez}},\ and\ \bibinfo {author} {\bibfnamefont {J.~M.}\ \bibnamefont
  {Gray}},\ }\bibfield  {title} {\enquote {\bibinfo {title} {Well-posed and
  ill-posed behaviour of the $\mu(i)$-rheology for granular flow},}\ }\href
  {https://doi.org/10.1017/jfm.2015.412} {\bibfield  {journal} {\bibinfo
  {journal} {J. Fluid Mech.}\ }\textbf {\bibinfo {volume} {779}},\ \bibinfo
  {pages} {794--818} (\bibinfo {year} {2015})}\BibitemShut {NoStop}%
\bibitem [{\citenamefont {Otsuki}\ and\ \citenamefont
  {Yoshii}(2025)}]{Otsuki2025}%
  \BibitemOpen
  \bibfield  {author} {\bibinfo {author} {\bibfnamefont {M.}~\bibnamefont
  {Otsuki}}\ and\ \bibinfo {author} {\bibfnamefont {K.}~\bibnamefont
  {Yoshii}},\ }\bibfield  {title} {\enquote {\bibinfo {title} {Scaling laws for
  flows of jammed frictionless granular materials between parallel plates},}\
  }in\ \href {https://doi.org/10.1051/epjconf/202534004002} {\emph {\bibinfo
  {booktitle} {EPJ Web Conf.}}},\ Vol.\ \bibinfo {volume} {340}\ (\bibinfo
  {publisher} {EDP Sciences},\ \bibinfo {year} {2025})\ p.\ \bibinfo {pages}
  {04002}\BibitemShut {NoStop}%
\bibitem [{\citenamefont {Haji-Sheikh}\ \emph {et~al.}(1984)\citenamefont
  {Haji-Sheikh}, \citenamefont {Lakshimanarayanan}, \citenamefont {Lou},\ and\
  \citenamefont {Ryan}}]{Haji-Sheikh1984}%
  \BibitemOpen
  \bibfield  {author} {\bibinfo {author} {\bibfnamefont {A.}~\bibnamefont
  {Haji-Sheikh}}, \bibinfo {author} {\bibfnamefont {R.}~\bibnamefont
  {Lakshimanarayanan}}, \bibinfo {author} {\bibfnamefont {D.~Y.~S.}\
  \bibnamefont {Lou}},\ and\ \bibinfo {author} {\bibfnamefont {P.}~\bibnamefont
  {Ryan}},\ }\bibfield  {title} {\enquote {\bibinfo {title} {{Confined Flow in
  a Partially-Filled Rotating Horizontal Cylinder}},}\ }\href
  {https://doi.org/10.1115/1.3243115} {\bibfield  {journal} {\bibinfo
  {journal} {J. Fluids Eng.}\ }\textbf {\bibinfo {volume} {106}},\ \bibinfo
  {pages} {270--278} (\bibinfo {year} {1984})}\BibitemShut {NoStop}%
\bibitem [{\citenamefont {Gray}(2001)}]{Gray2001}%
  \BibitemOpen
  \bibfield  {author} {\bibinfo {author} {\bibfnamefont {J.}~\bibnamefont
  {Gray}},\ }\bibfield  {title} {\enquote {\bibinfo {title} {Granular flow in
  partially filled slowly rotating drums},}\ }\href
  {https://doi.org/https://doi.org/10.1017/S0022112001004736} {\bibfield
  {journal} {\bibinfo  {journal} {J. Fluid Mech.}\ }\textbf {\bibinfo {volume}
  {441}},\ \bibinfo {pages} {1--29} (\bibinfo {year} {2001})}\BibitemShut
  {NoStop}%
\bibitem [{\citenamefont {Hirt}\ and\ \citenamefont
  {Nichols}(1981)}]{Hirt1981}%
  \BibitemOpen
  \bibfield  {author} {\bibinfo {author} {\bibfnamefont {C.}~\bibnamefont
  {Hirt}}\ and\ \bibinfo {author} {\bibfnamefont {B.}~\bibnamefont {Nichols}},\
  }\bibfield  {title} {\enquote {\bibinfo {title} {Volume of fluid (vof) method
  for the dynamics of free boundaries},}\ }\href
  {https://doi.org/https://doi.org/10.1016/0021-9991(81)90145-5} {\bibfield
  {journal} {\bibinfo  {journal} {J. Comput. Phys.}\ }\textbf {\bibinfo
  {volume} {39}},\ \bibinfo {pages} {201--225} (\bibinfo {year}
  {1981})}\BibitemShut {NoStop}%
\bibitem [{\citenamefont {Gueyffier}\ \emph {et~al.}(1999)\citenamefont
  {Gueyffier}, \citenamefont {Li}, \citenamefont {Nadim}, \citenamefont
  {Scardovelli},\ and\ \citenamefont {Zaleski}}]{Gueyffier1999}%
  \BibitemOpen
  \bibfield  {author} {\bibinfo {author} {\bibfnamefont {D.}~\bibnamefont
  {Gueyffier}}, \bibinfo {author} {\bibfnamefont {J.}~\bibnamefont {Li}},
  \bibinfo {author} {\bibfnamefont {A.}~\bibnamefont {Nadim}}, \bibinfo
  {author} {\bibfnamefont {R.}~\bibnamefont {Scardovelli}},\ and\ \bibinfo
  {author} {\bibfnamefont {S.}~\bibnamefont {Zaleski}},\ }\bibfield  {title}
  {\enquote {\bibinfo {title} {Volume-of-fluid interface tracking with smoothed
  surface stress methods for three-dimensional flows},}\ }\href
  {https://doi.org/https://doi.org/10.1006/jcph.1998.6168} {\bibfield
  {journal} {\bibinfo  {journal} {J. Comput. Phys.}\ }\textbf {\bibinfo
  {volume} {152}},\ \bibinfo {pages} {423--456} (\bibinfo {year}
  {1999})}\BibitemShut {NoStop}%
\bibitem [{\citenamefont {Li}, \citenamefont {Renardy},\ and\ \citenamefont
  {Renardy}(2000)}]{Li2000}%
  \BibitemOpen
  \bibfield  {author} {\bibinfo {author} {\bibfnamefont {J.}~\bibnamefont
  {Li}}, \bibinfo {author} {\bibfnamefont {Y.}~\bibnamefont {Renardy}},\ and\
  \bibinfo {author} {\bibfnamefont {M.}~\bibnamefont {Renardy}},\ }\bibfield
  {title} {\enquote {\bibinfo {title} {Numerical simulation of breakup of a
  viscous drop in simple shear flow through a volume-of-fluid method},}\ }\href
  {https://doi.org/10.1063/1.870305} {\bibfield  {journal} {\bibinfo  {journal}
  {Phys. Fluids}\ }\textbf {\bibinfo {volume} {12}},\ \bibinfo {pages}
  {269--282} (\bibinfo {year} {2000})}\BibitemShut {NoStop}%
\bibitem [{\citenamefont {Yokoi}(2007)}]{Yokoi2007}%
  \BibitemOpen
  \bibfield  {author} {\bibinfo {author} {\bibfnamefont {K.}~\bibnamefont
  {Yokoi}},\ }\bibfield  {title} {\enquote {\bibinfo {title} {{Efficient
  implementation of THINC skheme: A simple and practical smoothed VOF
  algorithm}},}\ }\href {https://doi.org/10.1016/j.jcp.2007.06.020} {\bibfield
  {journal} {\bibinfo  {journal} {J. Comput. Phys.}\ }\textbf {\bibinfo
  {volume} {226}},\ \bibinfo {pages} {1985--2002} (\bibinfo {year}
  {2007})}\BibitemShut {NoStop}%
\bibitem [{\citenamefont {Amsden}\ and\ \citenamefont
  {Harlow}(1970)}]{Amsden1970}%
  \BibitemOpen
  \bibfield  {author} {\bibinfo {author} {\bibfnamefont {A.}~\bibnamefont
  {Amsden}}\ and\ \bibinfo {author} {\bibfnamefont {F.}~\bibnamefont
  {Harlow}},\ }\bibfield  {title} {\enquote {\bibinfo {title} {A simplified mac
  technique for incompressible fluid flow calculations},}\ }\href
  {https://doi.org/https://doi.org/10.1016/0021-9991(70)90029-X} {\bibfield
  {journal} {\bibinfo  {journal} {J. Comput. Phys.}\ }\textbf {\bibinfo
  {volume} {6}},\ \bibinfo {pages} {322--325} (\bibinfo {year}
  {1970})}\BibitemShut {NoStop}%
\bibitem [{\citenamefont {Dodd}\ and\ \citenamefont
  {Ferrante}(2014)}]{Dodd2014}%
  \BibitemOpen
  \bibfield  {author} {\bibinfo {author} {\bibfnamefont {M.~S.}\ \bibnamefont
  {Dodd}}\ and\ \bibinfo {author} {\bibfnamefont {A.}~\bibnamefont
  {Ferrante}},\ }\bibfield  {title} {\enquote {\bibinfo {title} {A fast
  pressure-correction method for incompressible two-fluid flows},}\ }\href
  {https://doi.org/10.1016/j.jcp.2014.05.024} {\bibfield  {journal} {\bibinfo
  {journal} {J. Comput. Phys.}\ }\textbf {\bibinfo {volume} {273}},\ \bibinfo
  {pages} {416--434} (\bibinfo {year} {2014})}\BibitemShut {NoStop}%
\bibitem [{\citenamefont {Weymouth}\ and\ \citenamefont
  {Yue}(2011)}]{Weymouth2011}%
  \BibitemOpen
  \bibfield  {author} {\bibinfo {author} {\bibfnamefont {G.}~\bibnamefont
  {Weymouth}}\ and\ \bibinfo {author} {\bibfnamefont {D.}~\bibnamefont {Yue}},\
  }\bibfield  {title} {\enquote {\bibinfo {title} {Boundary data immersion
  method for cartesian-grid simulations of fluid-body interaction problems},}\
  }\href {https://doi.org/https://doi.org/10.1016/j.jcp.2011.04.022} {\bibfield
   {journal} {\bibinfo  {journal} {J. Comput. Phys.}\ }\textbf {\bibinfo
  {volume} {230}},\ \bibinfo {pages} {6233--6247} (\bibinfo {year}
  {2011})}\BibitemShut {NoStop}%
\bibitem [{\citenamefont {van~der Vorst}(1992)}]{Vorst1992}%
  \BibitemOpen
  \bibfield  {author} {\bibinfo {author} {\bibfnamefont {H.~A.}\ \bibnamefont
  {van~der Vorst}},\ }\bibfield  {title} {\enquote {\bibinfo {title}
  {{Bi-CGSTAB}: {A} fast and smoothly converging variant of {Bi-CG} for the
  solution of nonsymmetric linear systems},}\ }\href
  {https://doi.org/10.1137/0913035} {\bibfield  {journal} {\bibinfo  {journal}
  {SIAM J. Sci. Comput.}\ }\textbf {\bibinfo {volume} {13}},\ \bibinfo {pages}
  {631--644} (\bibinfo {year} {1992})}\BibitemShut {NoStop}%
\bibitem [{\citenamefont {Chauchat}\ and\ \citenamefont
  {M{\'e}dale}(2014)}]{Chauchat2014}%
  \BibitemOpen
  \bibfield  {author} {\bibinfo {author} {\bibfnamefont {J.}~\bibnamefont
  {Chauchat}}\ and\ \bibinfo {author} {\bibfnamefont {M.}~\bibnamefont
  {M{\'e}dale}},\ }\bibfield  {title} {\enquote {\bibinfo {title} {A
  three-dimensional numerical model for dense granular flows based on the
  $\mu(i)$ rheology},}\ }\href
  {https://doi.org/https://doi.org/10.1016/j.jcp.2013.09.004} {\bibfield
  {journal} {\bibinfo  {journal} {J. Comput. Phys.}\ }\textbf {\bibinfo
  {volume} {256}},\ \bibinfo {pages} {696--712} (\bibinfo {year}
  {2014})}\BibitemShut {NoStop}%
\bibitem [{\citenamefont {Lin}\ and\ \citenamefont {Yang}(2020)}]{Lin2020}%
  \BibitemOpen
  \bibfield  {author} {\bibinfo {author} {\bibfnamefont {C.}~\bibnamefont
  {Lin}}\ and\ \bibinfo {author} {\bibfnamefont {F.}~\bibnamefont {Yang}},\
  }\bibfield  {title} {\enquote {\bibinfo {title} {Continuum simulation for
  regularized non-local $\mu(i)$ model of dense granular flows},}\ }\href
  {https://doi.org/https://doi.org/10.1016/j.jcp.2020.109708} {\bibfield
  {journal} {\bibinfo  {journal} {J. Comput. Phys.}\ }\textbf {\bibinfo
  {volume} {420}},\ \bibinfo {pages} {109708} (\bibinfo {year}
  {2020})}\BibitemShut {NoStop}%
\bibitem [{\citenamefont {Lees}\ and\ \citenamefont
  {Edwards}(1972)}]{Lees1972}%
  \BibitemOpen
  \bibfield  {author} {\bibinfo {author} {\bibfnamefont {A.}~\bibnamefont
  {Lees}}\ and\ \bibinfo {author} {\bibfnamefont {S.}~\bibnamefont {Edwards}},\
  }\bibfield  {title} {\enquote {\bibinfo {title} {The computer study of
  transport processes under extreme conditions},}\ }\href
  {https://doi.org/10.1088/0022-3719/5/15/006} {\bibfield  {journal} {\bibinfo
  {journal} {J. Phys. C}\ }\textbf {\bibinfo {volume} {5}},\ \bibinfo {pages}
  {1921} (\bibinfo {year} {1972})}\BibitemShut {NoStop}%
\end{thebibliography}%

\end{document}